\documentclass[journal=jctc,manuscript=article]{achemso}
\SectionNumbersOn

\usepackage[T1]{fontenc} 

\usepackage{algorithm}
\usepackage{algpseudocode}
\usepackage{csquotes}
\usepackage{enumerate}
\usepackage[font=small,labelfont=bf]{caption}
\usepackage{booktabs}
\usepackage{subcaption}
\usepackage{wrapfig}
\usepackage{listings}
\usepackage{amsmath}

\let\oldAA\AA
\renewcommand{\AA}{\text{\textup{~\oldAA}}}

\usepackage{soul}
\usepackage{hyperref}
\usepackage{xcolor}
\hypersetup{
	colorlinks,
	linkcolor={red!50!black},
	citecolor={blue!50!black},
	urlcolor={blue!80!black}
}
\usepackage{cleveref}

\author{Luca Gagliardi}
\email{luca.gagliardi@iit.it}
\affiliation[IIT]
{Istituto Italiano di Tecnologia}
\author{Walter Rocchia}
\email{walter.rocchia@iit.it}
\affiliation[IIT]
{Istituto Italiano di Tecnologia}
\title[]
  {SiteFerret: beyond simple pocket identification in proteins
  }

\abbreviations{IR,NMR,UV}
\keywords{American Chemical Society, \LaTeX}

\begin{document}

\begin{abstract}
	\noindent We present a novel method for the automatic detection of pockets on protein molecular surfaces.
	The algorithm is  based on an ad hoc hierarchical clustering of virtual SES probe spheres obtained from the geometrical primitives generated by the NanoShaper software.
	The final ranking of putative pockets is based on the Isolation Forest method, an unsupervised learning approach originally developed for anomaly detection. A detailed importance analysis of pocket features provides insight on which geometrical (clustering) and chemical (residues) properties characterize a good binding site. The method also provides a segmentation of pockets into smaller subpockets. We prove that subpockets are a reliable representation that pinpoint the binding site with greater precision.
	Site Ferret is outstanding in its versatility, accurately predicting a wide range of binding sites, from small molecules to peptides and difficult shallow sites. 
\end{abstract}

\section{Introduction}

The automated prediction of protein active sites for biological processes (e.g.~protein-protein or protein-ligand binding regions) is a central problem in computational biophysics. It has fundamental implications for structural biology and drug design.
Predicting protein–ligand binding regions is particularly relevant in this respect. Indeed, identifying candidate binding sites is an essential preparatory step for rational drug design.

Due to its importance and difficulty, the problem of binding site prediction has been tackled by many methods.
These methods fall into three broad categories\cite{Simoes2017}:
a) Evolutionary and template-based (i.e. based on multiple sequence alignments)\cite{Brylinski2008}. These methods mainly address the problem from a biological perspective. 
b) Energy-based\cite{Ghersi2009,An2005}
(binding sites are assessed by estimating the interaction energies between protein atoms and a probe representing a small molecule). These methods address the problem from a physico-chemical standpoint.
c) Geometry-based. These methods use the properties of the molecular surface to infer potential binding sites. 
\\In this work, we focus on geometric methods.

\paragraph{Pocket generation in geometric methods.} Geometric methods fall into the following subcategories:
1.~Sphere-based: pockets and cavities are defined using spherical probes to fill voids on the surface\cite{Laskowski1995,Brandy1999}. 
2.~Grid based: the protein volume is mapped onto a 3D grid. Then, according to different geometric criteria, one identifies the grid points belonging to the pockets/cavities\cite{Huang2006,Weisel2007,Tripathi2010,Marchand2021}. 
3.~Tessellation/alpha-shape-based: these methods define pockets via filtered subcomplexes of the Delaunay triangulation of the molecular surface (alpha-shape)\cite{Edelsbrunner1998,LeGuilloux2009}.
4.~Surface-based: this much smaller set of methods identifies binding sites using local analytical geometric properties of the molecular surface (e.g. curvature)\cite{Dias2017}. There is renewed interest in using surface-based methods to address the less explored problem of protein-protein interaction\cite{Milanetti2021}.
Several approaches use a mixture of the above methods (e.g.~sphere and grid)\cite{Yu2010}.

In geometric methods, a special role is played by the molecular surface (MS) concept. 
A protein's MS is the region accessible by the solvent (water), and is the natural search area when seeking protein-ligand binding regions. Here, using the MS definition, we focus on the solvent-excluded surface (SES) or Connolly surface of a protein \cite{Richards1977,Connolly1983,Chen2010}, which is defined as the separation surface between locations that a spherical probe representing water can access, and locations where the probe is hindered by the protein's atoms, described as hard spheres endowed with Van der Waals radii. Geometrically, the SES can be imagined as the result of the spherical water probe rolling over the protein's Van der Waals surface. The SES comprises exposed regions that are locally convex and coincide with the atom parts that most protrude into the solvent, and concave reentrant regions. For more details on the SES and how it differs from the SAS (solvent-accessible surface), the reader is referred to the Supplementary Material of Ref.~\cite{Reis2022}
In this framework, one should seek a potential binding site among the concave regions of the MS (e.g. pockets, clefts/grooves, invaginations).

\paragraph{Ranking the candidates.} In most site identification methods, the second step is to rank the identified candidates.
Often, ranking criteria rely on physics-based scores, which are not trained on available data. Despite its simplicity, one of the most widely adopted and successful scoring systems of this kind is based on volume \cite{Macari2019,Gagliardi2022}. Alternative descriptors include the \emph{degree of buriedness}~\cite{Marchand2021} and descriptors based on chemical or biological properties, such as the evolutionary conservation of the constituting residues \cite{Huang2006,Glaser2004}.

Different approaches rank pockets according to more complex scoring functions that are trained or fitted on available datasets~\cite{LeGuilloux2009,Halgren2009,Hajduk2005}.
Such scores, obtained via standard techniques such as logistic regression~\cite{Schmidtke2010} or Support Vector Machine~\cite{Volkamer2012}, estimate the probability that a pocket is druggable.

Given the growing amount of crystallographic data, modern computational algorithms can leverage a large amount of information.
Therefore, data-driven/machine learning (ML) methods are now emerging\cite{Krivak2018,Macari2019,Gainza2020,Mylonas2021}. As described above, statistical learning methodologies are used in the scoring phase only. However, ML methods can also be used as a standalone tool to directly select the binding region of interest.
ML methods may heavily rely on chemical information that is gathered (learned) from large protein-ligand binding datasets containing labelled examples~\cite{Schmidtke2010,Krivak2018}. These often rely on some (strong) assumptions about the nature and distribution of negative samples (e.g.~regions not observed in contact with the ligand are usually labelled as nonbinding)~\cite{Jimenez2017}. This is an open question, since truly negative samples can hardly be ascertained, and may become positive as soon as a ligand is found to bind them. For a conceptualisation of this problem and a possible solution in the field of ML and kernel methods, the reader is referred to Ref.~\cite{Decherchi2022}.
Despite the increasing presence of standalone ML approaches, there is still interest in geometric methods to better rank, interpret, and hierarchically segment binding regions into subregions of interest (subpockets)\cite{Tripathi2010,Volkamer2012,Marchand2021,Simoes2017}.

\paragraph{State-of-the-art site predictors.}
 In this work, we extensively use Fpocket as a benchmark for binding site identification, and NanoShaper's pocket detection function, which we combine with a volume-based ranking. 

Fpocket is an open-source tool based on Voronoi tessellation and alpha-shape theory, which is widely used and serves as standard reference for benchmark in the literature\cite{LeGuilloux2009,Schmidtke2010}.

NanoShaper (NS)\cite{Decherchi2013,Decherchi2019} is an efficient software for building and triangulating complex manifold surface representations of the molecular surface (MS) according to several definitions, mainly the Connolly/SES. Although NS was mainly designed for construction and triangulation of MSs, it also offers a pocket detection function\cite{DecherchiBook}.
This function defines pockets as the volumetric difference between the space regions enclosed within the SESs of the protein obtained with two different probe radii, $3\AA$ and $1.4\AA$ (the water molecule's effective radius). The implementation is grid-based because it flags the grid points that are both inside the $3\AA$ SES and outside the $1.4\AA$ SES. Once these grid points are identified, a filtering procedure is adopted and the pocket is represented by building the MS of the union of water spheres ($1.4\AA$) centered on the pocket grid points. Only pockets with a volume greater than that of three water molecules are returned.
As originally presented in Ref.~\cite{Gagliardi2022}, the customized volume-ranked pockets are referred to as NS-Volume\cite{NSVolume_github}.

\paragraph{SiteFerret.}
In this work, we introduce SiteFerret, a new geometry-based approach that identifies good candidate binding sites by leveraging the information gathered from spherical probes' clustering events obtained from the construction of SESs at different probe radii. To create a valuable benchmark, we started from the freely available bindingMOAD dataset\cite{Benson2007}. For the first time, we rank our candidate pockets with the Isolation Forest algorithm, a classical anomaly detector method, allowing us to circumvent the two-class discrimination paradigm.
Below, we describe in detail the database used and the SiteFerret tool, and provide extensive comparisons against existing methods.

Specifically, we independently tested Fpocket and NS-Volume on the same datasets and with the same evaluation metrics used to test SiteFerret.
Indirect comparison with other methods are abundant for one of the used databases, namely LIGSITE-PocketPicker, where many methods including Fpocket have been tested. However, it is hard to perform a direct and fair comparison. This is due to its sensitivity to the specific criteria adopted to evaluate the ranking. With this in mind, we also provide qualitative comparisons with DogSite\cite{Volkamer2010} and CAVIAR\cite{Milanetti2021}, two geometry-based pocket detection methods that, like SiteFerret, return a pocket segmentation in terms of subpockets.
Interestingly, we also performed a direct comparison with a recent deep-neural-network-based method, DeepSurf\cite{Mylonas2021}.

\section{Material and Methods}

The algorithm proposed in this work is based on the SES constructed by the NanoShaper software\cite{Connolly1983,Decherchi2013}. The SES is constructed for different probe radii, ranging from $1.4\AA$ to $3.0\AA$. NanoShaper was instructed to report the probe spheres in the reentrant regions which are then clustered together in order to trace the concave regions of the SES. 
To avoid arbitrary assumptions about the distribution of negative samples and to consider this as a real \emph{one-class} classification problem, our scoring strategy uses an unsupervised Isolation Forest anomaly detection method\cite{Liu2008}. The classifier was trained on geometric features provided by the pocket generation step and on chemical information. 

Below, we start by illustrating the datasets, how the pockets are generated, and the metric used to evaluate how close a pocket is to an observed binding site.
Datasets and metrics form the ground truth against which SiteFerret is trained and assessed.
The third part of this section illustrates how the learning is performed by treating this as an \emph{anomaly detection} problem.
Finally, we discuss how the final ranking is obtained and presented, and how subpockets are considered, if present.

\subsection{The datasets}
\label{sec:database}

The main dataset is derived from the binding-MOAD database\cite{Benson2007} (BM).
The structures are selected by considering complexes with ligand molecular weights greater than 200~Da, a resolution better than 2\AA, available binding data, and limited redundancy (sequence identity $\leq 90\%$). 
The Binding MOAD separates ‘‘valid’’, i.e. biologically relevant, from ‘‘invalid’’ ligands (co-factors or due to the crystallization process). We only consider the former.
Note that, in contrast to what is often done, we include multimeric structures in order to also consider the sites at the interface between different monomers.

We also consider a second dataset, originally introduced to evaluate the LIGSITE$^{\text{csc}}$ method\cite{Huang2006}. It contains 48 complexes (holo) taken from the RCSB Protein Data Bank as well as their corresponding 48 unbound (apo) structures. It has been used as a benchmark in several studies\cite{Huang2006,Weisel2007,LeGuilloux2009,Volkamer2010}. We refer to this dataset as the LIGSITE-PocketPicker database (LP).

Finally, we create a further subset of the BM database by including only protein-peptide cases.

\paragraph{Data preparation}

The BM database was processed using a homemade Python script\cite{Gagliardi2022,lFetch_github} which: i) Establishes which subset of the PDB file represents the valid ligand according to the BM website information; ii) Removes the ligand and HETATM from the input PDB file and creates a PQR file using the AMBER force field via the pdb2pqr software\cite{Unni2011}; iii) Exports the valid ligands in an xyz file; iv) Creates a text lookup table mapping protein structure and ligand(s), discarding any invalid ligands and any moiety that has no full correspondence with what is expected from the MOAD naming scheme.
Furthermore, we also filtered out structures containing more than 10000 lines in the PQR format.
This initially resulted in a dataset of 1100 structures and 1808 ligand binding sites (the same structure can have more than one valid co-crystallized ligand). 
For multiple co-crystallized ligands, in order to avoid excessive overlap between the respective sites, we performed a further filtering. First, for each ligand we defined the corresponding binding region as the set of protein heavy atoms lying within $5\AA$ of any of the ligand's heavy atoms. Then, we evaluated the ratios of the cardinality of the intersection set between every pair of binding regions and that of the individual regions. If one was above $50\%$, the corresponding region was removed. This allowed us to exclude small sites, which are essentially included in larger ones. Otherwise, we evaluated the degree of overlap between the observed binding regions via their Jaccard index\footnote{Given two sets $\mathcal{A}$, $\mathcal{B}$, the Jaccard index is 
\begin{equation}
		J = \frac{|\mathcal{A}\cap \mathcal{B}|}{|\mathcal{A}\cup \mathcal{B}|}
\end{equation}
}. A Jaccard index greater than or equal to $30\%$ led us to discard the region with the higher ratio among those calculated in the previous step.

This filtering resulted in a final set of 1762 binding sites.
Finally, the peptide-binding sites were also removed, leaving a total of 1647 protein-small-molecule binding sites. 



 

As per the preparation of the LP database, for homogeneity and comparability with previous studies, we removed five holo structures from the LP database, as previously described (PDB codes: 1CDO, 5CNA, 1IGJ, 1SWB, 1A4J)\cite{Volkamer2010}. 
We also discarded one of the two co-crystallized ligands from the 4PHV structure. This is because it occupies exactly the same binding site as another ligand. After this pruning, the total number of remaining binding sites was 57.

Concerning the apo structures, we first localized
the reference \enquote{ground truth} binding sites 
by identifying the protein residues within $5\AA$ of any co-crystallized ligand heavy atom in the corresponding holo structures. Then, by sequence alignment, we found a matching of residues in the apo form. Of these, we kept only the residues with at least one \emph{solvent-exposed} atom (in the apo structure). The solvent-exposed residues were determined by considering the SES surface of the apo structure with probe radius $1.4\AA$\cite{solventExposed_github} (see \cref{sec:LP}, APO analysis). 

\subsection{Pocket generation}
\label{sec:algorithm}
In this approach, candidate pockets are first generated and then ranked. 
Similarly to other sphere-based\cite{Simoes2017} site detection methods, the main idea behind SiteFerret is that pockets, and cavities in general, can be well-approximated by the spherical probes used to build the SES in the concave/reentrant regions. In SiteFerret, these spheres are obtained from successive calls to NanoShaper, where the probe radius is gradually increased from that of a water molecule, $1.4\AA$, to a value of $3\AA$. These probe spheres are related to analytical geometric primitives representing reentrant concave patches in the SES surface\cite{Chen2010,Decherchi2013}. In the alpha-shapes-based construction procedure, the probe instances originating the concave patches of the SES correspond to the so-called "regular facet cells". Via NanoShaper, together with the probe instance position, we identify three protein atoms tangent to it and extract the corresponding trimming plane as well as the associated normal.
During the process of SES generation with a different probe radius, the probes are progressively clustered using a tailored hierarchical clustering algorithm. According to this algorithm, clustering events can occur \underline{only} between probes of equal radius or with radii differing by a single increment, corresponding to $\delta = 0.1$\AA. Given a probe of radius $r_i$, with $i$ being the $i-th$ call to NS, we will refer hereafter to \enquote{different} radii only when considering radii that differ by $\delta$ (i.e., $r_{i+1}= r_i+\delta$). 
Only specific conditions, summarized in \cref{code:cluster_tags}, allow a clustering event between two spheres.
Two of these conditions lead to events that are central features for a putative binding site: a \emph{bottleneck} and a so-called \emph{radial shift} event. 

\begin{table}
{
\floatname{algorithm}{Events causing probe clustering}
\begin{algorithm}[H]
	\renewcommand\thealgorithm{}
 \caption{}
	\begin{algorithmic}[1]
		\small
		\State Radial shift ($r_p\leftrightarrow r_p+\delta$): \textit{Terminal probe radius, depth index and direction (normal)} 
		\State Bottleneck ($r_p\leftrightarrow r_p$ and $r_p\leftrightarrow r_p+\delta$): \textit{Probe radius, normal to reference plane}
		\State Pyramidal aggregation ($r_p\leftrightarrow r_p+\delta$)
		\State Lateral aggregation ($r_p\leftrightarrow r_p$)
	\end{algorithmic}
\end{algorithm}
}

\caption{Clustering events \label{code:cluster_tags}}
\end{table}

Formally, a \emph{bottleneck} forms when two probes of equal or different radius are on the opposite sides of the same plane, and the projections of their centers on that plane coincide and fall within the triangle formed by the centers of the three tangent atoms defining the plane. 
More simply put, this type of event is observed in the presence of a narrowing of the SES, which can be approximately \enquote{sealed} on each side by two opposing probe spheres.

\begin{figure}[b!]
	\centering
	\caption{Sketch of some types of clustering event. 
	\label{fig:sketch}
	}
	\includegraphics[width=0.5\linewidth]{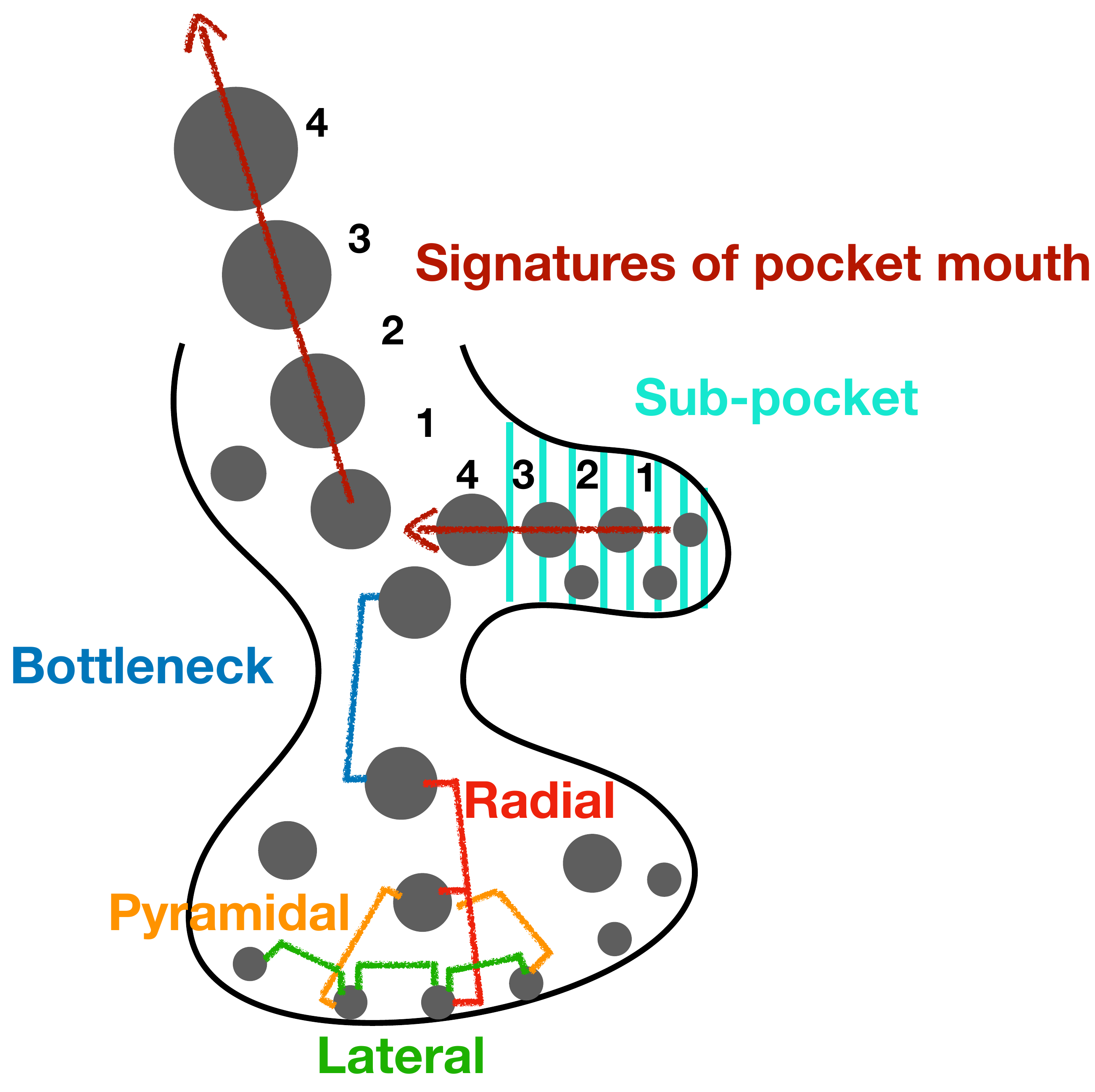}
\end{figure}

\emph{Radial shift} clustering events are related to the concept of pocket mouth.

The automated identification of mouth openings is one of the open problems in cavity detection methods. Our approach is based on the observation that, when the local surface features become insensitive to (large) probes, increasing the probe radius causes the probes to be aligned in the same direction, defined by the same protein atom triplet. We call these events 'radial shifts' and the number of aligned probes is the \emph{depth index}. 
The above observation explains the conventional definition of pockets in SiteFerret: a large cluster terminating in a series of aligned spherical probes.  
After several tests, we adopted for the radial shift a threshold value of 4 aligned spheres of increasing radius (which we define as \emph{significant alignment}). In addition, to enforce a minimum size, we also introduced a threshold value of at least $100$ clustered spheres. Slight variations in these two parameters (aligned "exit" spheres and minimum cluster size) do not significantly affect the results.
Thus, the concomitant realization of a radial shift of 4 aligned spheres (i.e. a \emph{a pseudo-mouth}) and a cluster size equal to or larger than 100 spheres defines a (conventional) pocket. Pockets can have more than one pseudo-mouth. 

It can happen that these pseudo-mouths are located in an intermediate position within a larger pocket. Therefore, one could provide an irreducible representation by recording all the minimal subclusters containing more than 100 spheres and presenting at least one pseudo-mouth. We call these subclusters \emph{subpockets}. 
Some types of clustering event and the conventional definition of mouths and of subpockets are illustrated in \cref{fig:sketch}. Interestingly, neither bottleneck nor radial shift clustering events require parametrization.

\begin{table}
{
\floatname{algorithm}{Parameters}
	\begin{algorithm}[H]
	\renewcommand\thealgorithm{}
		\caption{: 4 fixed, 2 tuned}
		\begin{algorithmic}[1]
			\small
			\State Terminal radius: $r^{\max} = 3\,$\AA 
			\State Probe increment: $\delta= 0.1\,$\AA
			\State 'Exit' alignment threshold: $4$
			\State Size threshold: $100$
			\State Maximum distance for \emph{lateral aggregation}: $\gamma$
			\State Maximum distance for \emph{pyramidal aggregation}: $\beta$
		\end{algorithmic}
	\end{algorithm}
}
\caption{Clustering parameters \label{code:cluster_params}}
\end{table}
Finally, the probe aggregation process considers two other types of clustering events: \emph{lateral} and \emph{pyramidal} aggregation.
The former is the clustering of probes with the same radius which are tangent to the same two out of three atoms and are on the same side of each of the tangent planes. The latter considers two distinct successive probe radii (therefore the probes are not aligned, and not tangent to the same three atoms, or it would fall into a \emph{radial shift} event). These two types of events without an extra constraint could lead to percolation of the "pocket" over the surface by successive aggregation of neighboring probes. Therefore, we introduce two parameters that control the maximum allowed distance for clustering two probes: $\gamma$ for \emph{lateral} and $\beta$ for \emph{pyramidal} aggregation. 
As shown in the supplementary material (Figures 3 and 4), the main effect of $\beta$ is to increase the total number of (putative) pockets (and subpockets). In turn, the effect of $\gamma$ is to decrease the total number of generated pockets while increasing their size. Based on this observation, an optimal trade-off could be found between the size and number of pockets. However, we found that this choice is suboptimal with respect to the ranking performance. 
Thus, $\beta$ and $\gamma$ are the basic hyperparameters of our method and have been determined by considering the algorithm's ranking performance. This is discussed more extensively in the following sections.
More details on the types of signature characterizing a clustering event are given in Table 1 of the Supplementary Material.

\paragraph{Subpockets and clustering: a comparison with similar methods.}
The concept of subpockets has been introduced by the CAVIAR\cite{Marchand2021} and DogSite\cite{Volkamer2010} methods in  different flavors. In CAVIAR, subpockets are the result of a post-processing step that borrows techniques and concepts from image recognition (watershed algorithm\cite{Beucher1994}).
In DogSite, similarly to our method, the concept of a subpocket is embedded in the procedure adopted to generate the putative pockets.
However, there are two main differences between SiteFerret and DogSite in this respect: i) We ensure that all subpockets are the smallest unique clusters fulfilling our definition of pocket. In DogSite, subpockets are large core elements, or cavities, which are subsequently merged into pockets. ii) Our concept of subpocket is independent from that of narrowing or bottleneck, but related to the concept of size and pocket-mouth/entrance, therefore we can have subpockets even in shallow sites.

Our approach is reminiscent of other sphere-based methods such as SURFNET or PASS\cite{Brandy1999,Laskowski1995}, but it differs from them as follows: i) SiteFerret's probe spheres are always related to the SES traced out by their corresponding radius (SURFNET uses non-SES-like spheres which are tangent to two atoms, while PASS uses SES-like spheres but grows them on the van Der Waals surface of previously placed spheres); ii) SiteFerret uses a (hierarchical) clustering approach; iii) SiteFerret's ranking score is trained (see the Ranking section).

In the sense of clustering, our method is reminiscent of Fpocket, which uses an ad hoc clustering of alpha spheres\footnote{Note that Fpocket uses different nonhierarchical clustering steps: 0. All alpha spheres are generated at once 1. single-linkage distance criterion, 2. center of mass linkage 3. linkage based on a majority vote.}. However, a fundamental difference with respect to Fpocket is that our spheres are probe instances and as such they are directly related to the protein's SES. Indeed, a drawback of Fpocket is that it can also include some protein atoms which are not solvent-exposed.

\begin{table}
{
\begin{algorithm}[H]
\renewcommand\thealgorithm{}
	\caption{Cluster generation }
	\begin{algorithmic}[1]
		\small
		\State NanoShaper call at probe radius $r_i$.
		\State Compute all pair distances between probes of current radius and sort.
		\State Cluster \emph{bottlenecks} among same radius probes.
		\State Cluster same radius probes: \emph{lateral aggregation}.
		\State NanoShaper call at probe radius $r_{i+1}= r_i+\delta$.
		\State Repeat $2\rightarrow5$.
		\State Compute all pair distances between probes of $r_i$ and $r_{i+1}$ and sort.
		\State Cluster \emph{bottlenecks} among probes of NS call $r_i$ and $r_{i+1}$.
		\State Cluster \emph{pyramidal aggregation} and \emph{radial shift} events.
		\State Update information of parent cluster event list using pre-order traversal of the cluster tree.
		\State If clusters comply with conventional pocket requirements, populate a putative pocket list.
		\State Repeat until the maximum probe radius is reached, $r_i=r_{\max}$ ($3\AA$).
    \State Filtering: \emph{subpockets} are identified.
	\end{algorithmic}
\end{algorithm}
}
\caption{Cluster generation steps. \label{code:algorithm}}
\end{table}

\subsection{Evaluation of the generated pockets}
\label{sec:ranking}

Several binding site prediction methods score generated pockets without adopting ML or statistical learning approaches. For instance, one simple yet successful score is based on volume. Other descriptors can be the degree of buriedness\cite{Marchand2021,Tripathi2010,Weisel2007,Hendlich1997} or residue conservation\cite{Huang2006}.

Some methods use ML techniques, such as regressing methods, trained on their datasets.
Fpocket, which is based on the alpha-shape complex of the Delaunay triangulation of protein atoms (see Tessellation Methods in the Introduction), returns a score based on five descriptors.
Other examples of linear regressors include the \enquote{Simple Score} of DogSiteScorer\cite{Volkamer2012}, the score of SiteMap\cite{Halgren2009}, 
or those derived in druggability studies, such as the hit rate obtained by NMR-based screening\cite{Hajduk2005}.  
A drawback of these types of approaches is that one must choose a (small) set of features on which to build the regression. On the one hand, this requires a good degree of a priori knowledge to select the most relevant parameters, which might vary considerably according to the system and problem considered. On the other hand, if one considers a large set of (probably redundant) descriptors, there is an increased risk of overfit and of losing generality. Another potential problem is to rely too much on the chosen score to be fitted. 

A slightly different approach is to directly machine learn a druggability/ligandability score from a dataset\cite{Schmidtke2010} according to a classification scheme\cite{Volkamer2012}.
These methods are very effective 
but, as with most standard ML approaches, they rely on the availability of datasets containing negative examples. This is a major issue in protein-ligand binding site recognition because the nature of the problem per se does not allow an unambiguous identification of a negative example\cite{Gagliardi2022,Decherchi2022}. While a few pockets have been labeled nondruggable sites as a result of pharmaceutical screening campaigns \cite{Cheng2007,Hajduk2005}, one must note that: i) nothing excludes the possibility that a drug may eventually be discovered for these pockets; ii) the concept of druggability, which is restricted to pharmaceutical applications, is not equivalent to ligandability. For example. to populate the negative dataset in the DogSiteScorer algorithm\cite{Volkamer2012}, so-called decoys are added to the dataset. These consist of putative pockets generated by the algorithm on regions of the structures where no ligand has been observed. This choice, however, is conceptually unsatisfactory since: i) it is method-specific (every method will return differently shaped putative pockets); ii) false negatives cannot be ruled out\cite{Amaro2019}.
Moreover, the ligand-binding process may structurally rearrange the binding region, further complicating definite claims over the nature of a binding site\cite{Surade2012}.

More recently, researchers have proposed a new family of data-driven approaches that use a different strategy. 
Rather than using ML to classify/score previously generated putative pockets, they directly predict the probability that a point on the surface belongs to a binding region. They then build putative binding regions by clustering nearby points with a predicted probability greater than some threshold. 
State-of-the-art data-driven ML approaches based on this procedure include P2Rank\cite{Krivak2018}, based on a Random Forest classifier, DeepSite\cite{Jimenez2017}, and DeepSurf\cite{Mylonas2021}, which use deep neural networks. Their performance is extremely good, despite the conceptual flaws described above\cite{Gagliardi2022}. Indeed, these methods intrinsically need negative examples (e.g.~in DeepSurf, surface points that are not observed binding are labeled as nonbinding sites) and suffer from data imbalance problems (the ratio between positive and negative labels is strongly uneven). 
Since the prediction of putative binding regions does not rely on a previous pocket generation phase, a strong advantage of this type of method is that they return a rather small set of putative pockets compared to more standard approaches.

After considering this rich and diverse set of approaches for evaluating the putative pockets, we opted for an unsupervised method that does not require explicit negatively labeled samples: the Isolation Forest anomaly detector.
Below, we identify a fairly sizeable number of features to describe pockets, mostly taken from the geometric clustering process, and we integrate them with others that are chemical in nature.

\subsubsection{Geometric and clustering features}

\noindent Given a pocket, we consider the ensemble of clustering events that generated the final cluster during its growth (child nodes). They are gathered by performing a "pre-order traversal" of the cluster (step 10 in the algorithm table).
\paragraph{Geometric features}\mbox{}\\ 
1. \emph{Number of entrances}.  
As a post-processing step, pocket entrances are obtained by considering
\emph{significantly aligned} probes (above the threshold of 4 aligned spheres) 
and grouping terminal probe spheres with a radius greater than $2.4\AA$. This clustering step is based on a standard single-linkage procedure where probes are clustered if their squared distance is smaller than the sum of their squared radii (orthogonality condition of the power distance\cite{Edelsbrunner1992}). 
Entrances are characterized by a center, an effective radius, and an average \emph{depth index}. The effective radius is defined as the average distance of all centers of the clustered entrance spheres with the geometric center of the cluster plus the average radius of the cluster. The average depth index is obtained by averaging all depth indexes of the clustered terminal spheres. 
If there are no significant alignments with a radius above the threshold, 
the (2.) \emph{buried} Boolean flag is assigned, meaning that the putative pocket is likely in a deep hollow.
Note that, by definition, subpockets have a single entrance associated with their unique pseudo-mouth. Again, if the terminal sphere is below $2.5\AA$, the subpocket is considered buried.

3. \emph{Average entrance (effective) radius}: The average of all entrance radii, where each entrance radius is the effective radius defined above.

4. \emph{(average) Entrance depth score}: Weighted average of the effective depth of entrances in the putative pocket. The weight is given by the number of spheres in the entrance cluster. This descriptor is normalized by the \emph{significant alignment} (fixed to 4, see in the previous section) so that 1 would correspond to the minimum requirement of a pocket pseudo-mouth, and greater than 1 signifies that the entrance is deep. 

5. \emph{Number of bottlenecks}\footnote{The same probe can also describes multiple bottleneck events when it is clustered to another probe of the same radius and with one or more larger radius probes (with different tangent atoms/planes). This situation describes, for instance, a superficial narrowing (canyon) that can be accessed from multiple directions.}.

6. \emph{Average bottlenecks radius}

7. \emph{Average large-aggregation radius}: Average radius of the large-aggregation clustering events defined in the following paragraph. The average is weighted by the number of elements in the clusters prior to merging. A large radius indicates that child node clusters are merged closer to superficial parts of the SES. 

8. \emph{Volume}: Given that clusters (putative pockets) are represented as an ensemble of overlapping probe spheres, a cluster is associated with a volume and a surface area. 
Volume and area are computed via the NanoShaper molecular surface triangulation of the pocket-clustered spheres (using the NS \enquote{Skin} surface option\cite{Decherchi2013}, which is more robust for a small set of strongly overlapping spheres)\footnote{An alternative analytical method\cite{Busa2005,Busa2012} was implemented (in Ref.\cite{arvoPY_github} a standalone Python interface), but it is significantly slower due to the high number of overlapping spheres in a cluster.}. The advantage of this approach is that it is fast and simultaneously produces the triangulation of the pocket mold, which is one of the outputs returned to the user.

\paragraph{Clustering-event-based features}\mbox{}\\
One of the novelties of our approach is that the clustering events used to build pockets are leveraged as descriptors for the following learning stage.

The following descriptors are heuristic scores related to the number of significant clustering events. 
We introduce an \emph{aggregation} list linked to the lateral and pyramidal clustering events. Given a cluster, the \emph{aggregation} list is a container where each entry contains further information about a clustering event (radius of the probe, cardinality of the clusters before merging) except for \emph{bottlenecks} and \emph{radial shifts}.
Similarly to the \emph{aggregation} list, given a cluster, we also define a \emph{persistence} list, which is a container where each entry is the \emph{depth index} (number of aligned spheres) of the sets of aligned spheres whose depth index is larger than the \emph{significant alignment} threshold. 
Below, some scores are normalized by the cluster number of elements to allow comparability of differently sized clusters.

9. \emph{Size}: Simple count of the cluster elements. This correlates to the volume but includes overlaps among spheres.

10. \emph{Aggregation score}: The normalized length of the \emph{aggregation} vector.

11. \emph{Persistence score}: The normalized sum of all \emph{persistence} entries. This corresponds to the number of aligned spheres (above threshold).

12. \emph{Clustering score}: The sum of aggregation and persistence scores. This is a measure of the total number of nontrivial clustering events. 

13. \emph{Large-aggregation score}: Similar to the \emph{aggregation score}, but considering a subset of the aggregation vector, which includes only major clustering events (i.e. only pyramidal and lateral aggregation events). These are selected by considering the relative size of two clusters before merging, which must be at least in a ratio of 1 to 5. That is, the smallest cluster before merging is at least one fifth the size of the largest one. To avoid including irrelevant events related to the early stages of the clustering process, we consider only events involving clusters larger than 10 probe spheres prior to merging.

\paragraph{Compactness}\mbox{}\\
The idea of correlating compactness with druggability is not new\cite{Hajduk2005,Krasowski2011}.
We here consider more classical evaluations and some ad hoc descriptors inspired by our specific clustering algorithm.

14. \emph{Volume ratio}: Ratio of volume to number of elements (spheres) in the cluster. This number correlates to the cluster's degree of compactness. A small value indicates that the spheres are more overlapping and vice versa. This is similar to the ratio between volume and area proposed in Ref.\cite{Hajduk2005}.

15. \emph{HW index}: (Hakon Wadell) Sphericity, a measure of how spherical an object is. This is given by the surface area of a sphere with an identical volume, divided by the object's actual surface area
\begin{equation}
	HW = \frac{\pi^{\frac{1}{3}}(6V)^{\frac{2}{3}}}{A}\, ,
\end{equation}
with A and V the volume of the putative pocket. The closer to 1 this index is, the more spherical is the cluster. This descriptor was proposed in Ref.\cite{Krasowski2011}.

The following are ad hoc heuristic compactness descriptors derived from the clustering process. 
Given the \emph{persistence} list detailed above, we introduce the following descriptors:

16. \emph{Ramification}: The standard deviation of the \emph{persistence} list. This is heuristically linked to the degree of deviation with respect to the average channeling (represented by the average significant-depth or persistence), and represents a measure of the cluster's complexity.

17. \emph{Protrusion}: The difference between the \emph{entrance} depth and the average \emph{persistence}, minus the ramification. The idea behind this descriptor is to obtain a measure of the cluster's complexity closer to the SES surface.

\subsubsection{Chemical features}
We can easily assign to each surface region the list of constituting atoms because they are tangent to corresponding probe spheres. Thus, we can also generate chemical features for every pocket. 
To infer the chemical features, we consider the percentages of representation of each of the 20 amino acids in a pocket's composition. A protein's residue is considered represented if at least one atom of that residue is tangent to any of the cluster probe spheres which comprise the pocket itself. The count is normalized by the total number of residues found in the pocket. A detailed statistical analysis of amino acid presence in binding pockets is shown in the Supplementary Material and thoroughly discussed in Ref.\cite{Khazanov2013}.
The other two chemical descriptors consider the global degree of hydrophobicity and hydrophilicity of each putative pocket, referred to as \emph{hydrophobic score} and \emph{hydrophilic score}, respectively.
Similarly to Fpocket, we consider the degree of hydrophobicity of the residues\cite{LeGuilloux2009,Monera1995},
dividing them into two broad families: hydrophilic and hydrophobic (following the table published at \url{https://gilles-hunault.leria-info.univ-angers.fr/Idas/proprietes.htm}, and reproduced in the Supplementary Material). The hydrophilic and hydrophobic scores are then computed by simply counting the number of residues in each class, normalized by the total number of residues. We exclude glycine from these counters due to its neutral behavior (although it participates in the normalization). 

As discussed in detail in the Supplementary Material, local hydrophilic and hydrophobic scores, accounting for the spatial distribution of the hydrophobic and hydrophilic residues within the pocket, were also implemented to identify if the binding pocket contains local parts that are rather hydrophobic (hydrophilic). However, this descriptor is not currently implemented because it did not improve performance and was computationally demanding (since it requires the computation of all pair distances between protein atoms of a putative pocket).

\begin{algorithm}
	\caption{Feature extraction and ranking \label{code:algorithm_post}}
	\begin{algorithmic}[1]
		\small
		\State Compute and extract all geometric (clustering) and chemical features of putative pockets with appropriate normalization.
		\State Load 2 couples (geometry and chemistry) of pretrained Isolation Forests (IFs): one trained on the \enquote{large} pockets, and one trained on the \enquote{small} (sub)pockets. The former is the main score, while the latter is used only to compare subpockets with each other.
		\State Compute the anomaly score from the (main) geometric and chemical IFs. Rank all putative pockets according to the average score.
		\State If any, rank the subpockets within each pocket. 
		\State Return to the user the pockets ranked according to point 3, and provide the subrank of subpockets, if any, according to point 4.
	\end{algorithmic}
\end{algorithm}

\subsubsection{Evaluation of the matching of the pockets and binding sites}
\label{sec:metrics}

To choose which pockets approximate the observed binding sites well enough to be fed into the ML tool and to assess the method's performance, we need to define an evaluation criterion. We required that our metric be based on the atom composition of putative and observed sites rather than on concepts that depend on the specific pocket construction or representation  (e.g.~spheres, alpha-spheres\cite{LeGuilloux2009}, grid-points\cite{Huang2006,Weisel2007,Volkamer2010,Tripathi2010,Marchand2021}). Since our method uses probe spheres that contact at least three atoms, we used the list of contacted atoms to assess the degree to which a pocket matches an observed binding site.

Many site detection methods consider a pocket to be a correct match if its center of mass lies within $4\AA$ of any ligand atom\cite{,LeGuilloux2009,Krivak2018}. In agreement with the initial suggestion of the Fpocket  authors\cite{LeGuilloux2009} (who define a mutual overlap criterion) and with the dogSite method\cite{Volkamer2010}, we instead adopted a score based on combining two figures of merit, as
detailed below. This metric was extensively tested in Ref.\cite{Gagliardi2022}.

\paragraph{Ligand coverage score (LC)} If we define 'contact' as the property of being closer than $5\AA$, LC represents the ratio of ligand heavy atoms that are in contact with at least one heavy atom of the putative pocket, divided by those in contact with at least one heavy atom of the entire protein.
For a given pocket, we indicate with $d(i,j)$ the Euclidean distance between the centers of two atoms, with $n_L$ being the number of ligand heavy atoms within 5\AA~of the protein that comprises the $\mathcal{L}$ set, and with $n_P$ being the number of protein heavy atoms that belong to the pocket, producing the $\mathcal{P}$ set, we have:
\begin{equation}
	\label{eq:lc}
	\mathrm{LC} = \frac{1}{n_{L}}
	\sum_{j\in\mathcal{L}}^{n_L}\delta_j,
	\quad 
	\delta_j =  
	\begin{cases}
		1\quad\text{if } \exists~ i\in\mathcal{P} : d(i,j)\leq 5\AA\\
		0 \quad \text{otherwise}\\
	\end{cases}
\end{equation}
An LC value close to 1, or $100\%$, denotes a pocket in contact with most of the cocrystallized ligand. 

\paragraph{Pocket coverage score (PC)} A good LC score alone does not exclude that there is a large part of the putative pocket that is not in contact with the ligand (for instance, the entire protein by definition always scores $100\%$ on LC).
To account for this possibility, we introduce the PC, which represents the fraction of protein surface atoms that belong to a pocket and that are within $5\AA$ of any heavy atom of the ligand. This score is the symmetric version of the LC:
\begin{equation}
	\label{eq:pc}
	\mathrm{PC} = \frac{1}{n_{P}}\sum_{i\in\mathcal{P}}^{n_P}\delta_i,
	\quad
	\delta_i =  
	\begin{cases}
		1\quad\text{if } \exists~j \in\mathcal{L} : d(i,j)\leq 5\AA\\
		0 \quad \text{otherwise} \\
	\end{cases}
\end{equation}
A large pocket coverage score implies that only a few atoms of the putative pocket are not in contact with the ligand. Again, this score alone would not be sufficient to correctly evaluate a prediction. Indeed, small pockets in contact with a large ligand would score very high in PC but could nevertheless miss a large portion of the binding region.

In summary, we consider a putative pocket to be a correct match if it scores at least $50\%$ in Ligand Coverage and at least $20\%$ in Pocket Coverage. The obtained evaluation scores are rounded up to the first decimal in percentage. In some cases, we made the requirements for the PC score stricter in order to see whether an approach can more precisely spot a binding site.

\subsubsection{Nonbinding pockets as anomalies: the Isolation Forest approach}
\label{sec:IsolationForest}

Due to the above-discussed issues inherent in defining negative samples for binding sites, we approached the problem of scoring putative pockets as a one-class discrimination problem\cite{Decherchi2022}. Namely, we assumed that we only hold samples of the positive class (also referred to as the normal or inlier class).
One-class learning is a task that typically arises in
outlier (anomaly) detection or, more generally, in discrimination data
mining problems, where it is too expensive or challenging to obtain examples of "the other" class\cite{Gagliardi2022,Itani2020,Decherchi2017}.

\begin{figure}[t]
	\centering
	\caption{Illustration of the procedure adopted by the Isolation Tree. Left panel, $d=2$ features. Normal point (inlier), blue. Anomaly, red. Right panel: sketch of the corresponding tree. Each branching corresponds to a split in the feature space (horizontal and vertical lines, in the 2D example in the left panel).
	Reworking of image in Refs.\cite{Hariri2021} and \cite{Liu2008}.
	\label{fig:IF}}
	\includegraphics[width=0.9\linewidth]{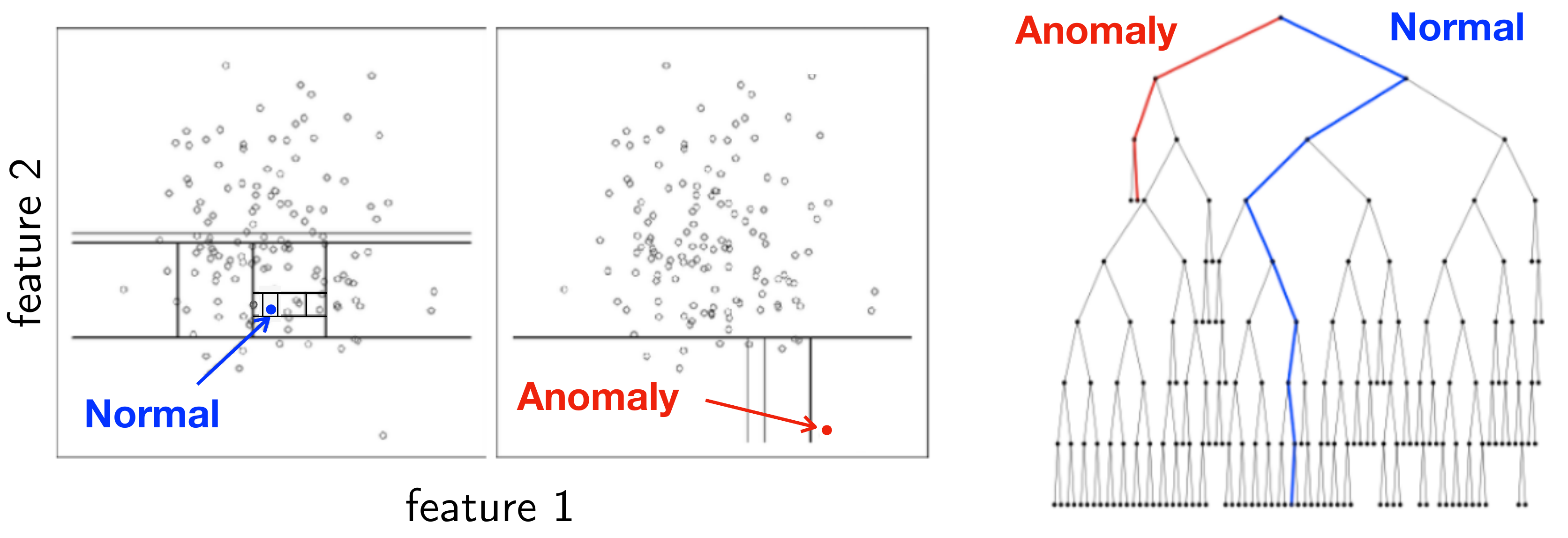}
\end{figure}

For this purpose, we borrowed a standard unsupervised anomaly detector method, the Isolation Forest (IF)\cite{Liu2008}.
The advantages of IF here are: 
\begin{itemize}
	\item it requires minimal parametrization;
	\item it can handle high-dimensional problems with a large number of irrelevant or redundant attributes (descriptors), which is ideal, considering our 17 geometric/clustering and 22 chemical descriptors;
	\item it is highly efficient, which makes it ideal for large datasets.
\end{itemize}
The central rationale of IF is that anomalies are more subject to isolation under random data partitioning with respect to normal points (see left panel of \cref{fig:IF}).
An IF comprises multiple binary decision trees (iTree) trained on different random subsets (subsampling) of the training dataset. In the training phase, each iTree is grown by dividing data at each node by randomly selecting an attribute (feature) and a split value. The process terminates when the tree reaches a height limit or each external node (leaf) contains a single observation. 
Given a sample $\mathbf{x}=(\mathbf{x}_1,\dots \mathbf{x}_n)$ with $\mathbf{x}_i$ the $i$-th observation, $n$ the subsample size, and $\mathbf{x}\in \mathcal{R}^d$ an observation vector in a d-dimensional feature space, 
we define $h(\mathbf{x})$ as the path length of an observation (pocket) measured as the number of edges traversed in a tree (right panel of \cref{fig:IF}).
Given a collection of trees, $E\left(h(\mathbf{x})\right)$ is the mean $h(\mathbf{x})$.
The anomaly score is defined as 
\begin{equation}
	s(\mathbf{x},n) = 2^{-\frac{E\left(h(\mathbf{x})\right)}{c(n)}}
\end{equation}
where $c(n)$ 
is a normalization factor corresponding to the average path length of an unsuccessful search in the Binary Search Tree.
Therefore, $s\rightarrow 1$ corresponds to a strong anomaly, and $s\rightarrow 0$ corresponds to a "normal" sample.

An improved version of IF, named Extended Isolation Forest, was recently proposed to address the issues related to assigning an anomaly score to data\cite{Hariri2021}.
The new version did not significantly improve our problem and it carried a significantly higher computational cost in terms of memory (about 10 times larger than for standard IF).


\subsubsection{Training, validation, and test}

The training set for the IF comprises about $86\%$ of the BM dataset, and amounts to a total of 1416 binding sites\footnote{Note that, for discrimination tasks/anomaly detection, the usual distinction between training and test set is blurry. Indeed, the training considers exclusively positive examples, while, in the test phase, all generated pockets are fed into the IF to be scored.}. The positive samples were selected by selecting all the \enquote{hitting} pockets according to the metrics detailed in \cref{sec:metrics}.
At this stage, the algorithm still contains 2 free hyperparameters, $\gamma$ and $\beta$.
We ran the clustering algorithm on the range $\gamma = 0 \to 0.7$ $\beta = 0 \to 0.6$. As illustrated in the Supplementary Material, this range covers many pocket numbers and volumes, and returns very high \enquote{hit} rates.
To maximize the training set, we used all the obtained positive examples for the considered range of hyperparameters. We also considered hitting subpockets as distinct positive samples. 
Since the Ligand Coverage score was on average higher than the Pocket Coverage score, we artificially replicated entries with a frequency proportional to the PC score, \cref{eq:pc}. This informs the IF about the best training samples by increasing the number of samples with a large PC score and thus biasing the scoring towards preferring smaller pockets.
The IF was implemented using the open-source Python library scikit-learn\cite{Pedregosa2011}. After several tests, the IF was trained with the following parameters: no bootstrap (i.e.~no random repetition of data across the different iTrees of the forest, since we already introduced replicas in the sample following the PC score), 
subsampling size of 256, and 10000 iTrees.
We chose to consider the geometric and chemical descriptors separately in two distinct IFs. This allowed us to gain insight into the discriminating power of geometry/clustering-based and chemistry-based ranking individually.
We found that, when the training included both \enquote{master} pockets (which contain one or more subpockets) and small pockets (such as subpockets), this negatively impacted the overall performance. This is probably due to their very different geometric signatures (see Supplementary Material). We therefore considered two distinct populations separately: "large" pockets and "small" pockets, where large pockets are master pockets that contain more than one subpocket 
 \footnote{The second group (small pockets) greatly outnumbers the first: 174224 against 32688 (considering the pockets generated with all 56 combinations of clustering parameters, $\beta$ and $\gamma$). }.
We therefore trained a total of 4 IFs (1 geometric + 1 chemical for the positive samples of each population).
In the validation stage, we selected the optimal clustering parameters, $\gamma$ and $\beta$, and identifed the combination of the 4 trained IFs delivering the best performance. 
In the training phase, we were only feeding the IFs with the positive samples. During validation, in contrast, we provided the already trained IFs with all the generated pockets, regardless of whether they hit an observed binding site or not. We then evaluated which combination of IFs, especially of hyperparameters, yielded the best results.

After several trials, we decided to use the chemical and geometric forests trained on the \enquote{large pockets} population to provide the main scoring for both large and small pockets.  
However, the IFs trained on the \enquote{small pockets} were used to rank subpockets within their parent master pocket, when present.
As illustrated in more detail in the Supplementary Material, we also found that the best ranking performance was achieved by averaging the score of the geometric and chemical IFs.

Finally, we found that $\gamma=0$ and $\beta=0.9$ gives the combination of clustering parameters that yields the best ranking. More details on the effect of clustering parameters are discussed in the Supplementary Materials.\\  
 As is customary, we assessed the method in the deployment (test) phase without free parameters and on a previously unseen set of putative sites.
 The main test set corresponded to approximately a fraction of $14\%$ of the entire BM dataset described in \cref{sec:database}, and it was similar to the test set first used in the Shrec2022 Computer Graphics benchmark\cite{Gagliardi2022}, with the only difference being that the 20 peptide-protein structures were not included. This resulted in a set of 229 protein-ligand pairs distributed on 150 structures (versus the 249 sites in Ref.\cite{Gagliardi2022}). 
 We then used the LP database, which is divided into bound (holo) and unbound (apo) structures. Finally, we performed a dedicated assessment of the protein-peptide structures found in the BM (115 entries, see \cref{sec:results:peptides}, which were not used in the training).

\paragraph{Feature importance analysis}

Given an unsupervised ML method, which is not constructed around a specifically designed scoring measure, it is particularly interesting to try to interpret the results by estimating how the different features impact the performance. For a simple ML model, such as linear or logistic regression, one can quickly evaluate the feature importance by analysing the coefficients associated with each feature. However, it is more difficult to interpret complex methods such as random forest or artificial neural networks.
Here, we adopted the Shapley Additive exPlanations (SHAP) framework\cite{Lundberg2020}, based on cooperative game theory, to estimate each feature's relative contribution to the model outcome. In particular, we used the treeSHAP algorithm~\cite{Lundberg2020}, which is specifically optimized for ensemble-based decision tree methods and thus compatible with IF anomaly detection\cite{Kim2021}.

\subsection{Ranking protocols and comparison between different methods}
\label{rank_protocols}
The problem of evaluating pocket detection algorithms in a reproducible and comparable way is far from trivial. Assessments of existing methods vary in both the ranking scheme adopted and the definition of a “correct prediction” (i.e. evaluation metrics). Here, similarly to what was done in Refs.\cite{Chen2011,Krivak2018,Mylonas2021}, we propose that, for a given structure with one or more known co-crystallized
ligands and a given method returning an ordered list of putative sites,
the final ranking position is given by the number of nonmatching pockets that occupy a higher position in the prediction list than the hit pocket. The normalization is then given by the number of observed binding sites (structure–ligand pairs) rather than the number of examined structures. In this way, we resolve the issue of structures having more than one crystallized ligand.

As anticipated in the introduction, we explicitly compared SiteFerret's performance with that of Fpocket and with a customized version of the NanoShaper pocket detection method, which we called NS-Volume\cite{NSVolume_github} and which ranks the pockets generated with NS by volume. 
These alternative tools were chosen to allow a fair comparison. Both Fpocket and NS-Volume could be run on the same datasets and with the very same evaluation metrics and ranking protocols because they also return the atoms associated with the generated pockets. For Fpocket, these are the atoms contacted by the alpha spheres. For NS, these are the surface atoms facing the triangulated volume which defines the pocket. This information is needed to compute the LC and PC scores in \cref{sec:metrics}.
Finally, we also compared SiteFerret to DeepSurf, a recent deep-neural-network-based method, on the test set of the BM dataset. This was because we could use the (partially unpublished) data from Ref.\cite{Gagliardi2022}, which used the same evaluation metrics as for SiteFerret.

\subsubsection{Master pockets and subpockets}

The raw output of SiteFerret is a score for the pockets found in the structures that are fed as input. As discussed above, our pocket generation method means that a generated pocket may contain zero, one, or several subpockets. If it contains one or more subpockets, we call it a "master pocket". As discussed in detail in the next section, for each returned master pocket, the user is provided with the ranking of internal subpockets and can inspect them.
The full list of generated pockets, which includes the subpockets isolated from the parent master pockets, is passed to the IFs for the scoring phase. While the identification of subpockets is interesting and in agreement with findings in some recent literature, it also poses the problem of whether the same scoring system can be used for both classes.
We tried to address this issue in the following ranking protocol. 

First, for each master pocket, we considered only its three top-ranking subpockets and discarded any further ones. Then, if a subpocket was ranked after its parent master pocket, it was skipped, since it was already contained in a pocket higher in ranking.
Unless explicitly specified, we did not consider the segmentation into subpockets when evaluating the master pocket. However, we tracked the presence of successful subpockets (according to the same evaluation metrics used for the parent pocket), as indicated in the footnotes of the tables.
We adopted two possible schemes when evaluating subpockets.

\paragraph{Single-Pocket Evaluation.} According to this protocol, if present, the three top-ranking subpockets replace their parent master pocket in the prediction chart. We empirically established that the master pocket's score (given by the IF trained on the population of "large" pockets) leads to the best ranking performance. Therefore, to build a ranking of the new ensemble of pockets, we decided that the score of a (former) subpocket is given by a penalized version of its parent master pocket's score, calculated proportionally to its subranking within the parent pocket: 
\begin{equation}
	\label{eq:subRanking}
	\tilde{s} = s\left(1+0.05\times (R-1)\right)
\end{equation} 
with $R\geq 1$ being the subpocket (integer) ranking position and $s$ being the master pocket's score (which is lower for higher ranking positions). Therefore, if a pocket has only one subpocket ($R=1$), it is simply substituted for the latter with no effect on its ranking score (see \cref{sec:IsolationForest}). When a pocket contains more than one subpocket, these are inserted into the ranking according to scores given by \cref{eq:subRanking}. Pockets with no subpockets are not affected.

\paragraph{Nested Evaluation.} In this protocol, if one of the three top-ranking subpockets of a master pocket matches a binding site according to the current evaluation metric (i.e. the combination of PC and LC scores), we label the parent master pocket as successful (even if the entire master pocket would not meet the PC requirement). When reported, LC and PC scores refer to the system i.e. the master pocket or, when it does not meet the coverage requirements, its subpocket, which match the binding site.

The Nested Evaluation concept stems from evidence that binding sites are often found as part of more elaborated structures, as observed elsewhere \cite{Marchand2021}. However, Nested Evaluation can also be seen as a way to implicitly increase the number of potential candidates, expecting the final user to look at more than 10 pockets. This is why we report the results from both protocols.

\subsection{Summary of SiteFerret output}

\begin{figure}
	\centering
	\caption{Examples of some of the visual outputs provided by SiteFerret on PDB code: 7TAA. a) SES of the structure (obtained via NanoShaper) and cocrystallized ligand shown together with the returned binding pocket, here the second top-ranked. The pocket is represented by the clustered probe spheres (transparent white), with entrance normal (red) and bottleneck normal (green). In orange, the protein atoms tangent to the pocket (artificially enlarged for visual purposes). b) An alternative representation of the pocket as the SES of the clustered spheres, also showing the corresponding tangent protein atoms.
		\label{fig:output}}
	\begin{subfigure}{0.45\linewidth}
		\includegraphics[width=\linewidth]{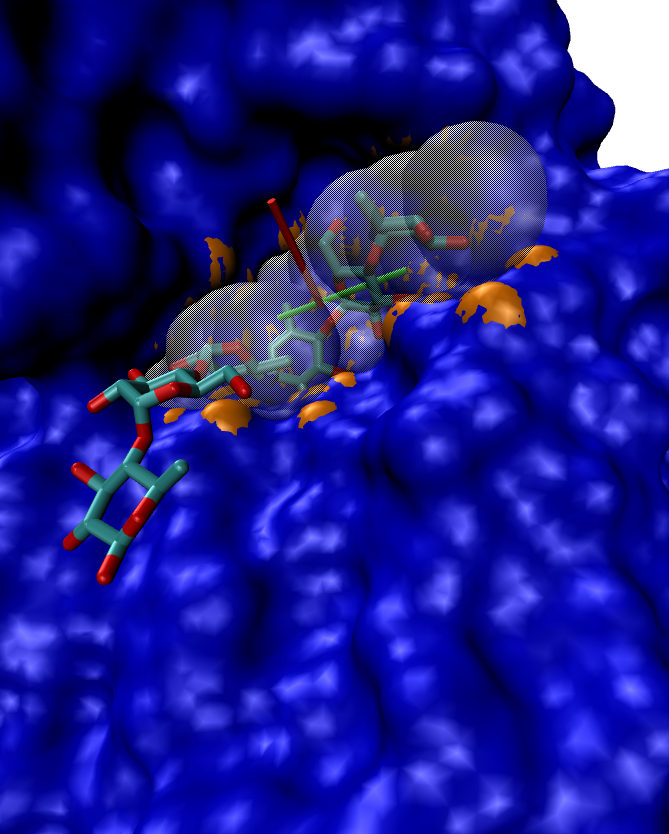}
	\end{subfigure}
	\begin{subfigure}{0.45\linewidth}
		\includegraphics[width=\linewidth]{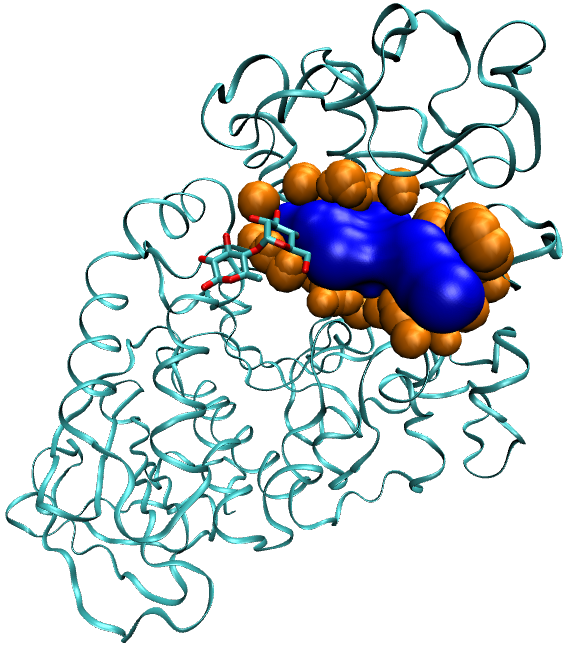}
	\end{subfigure}
\end{figure}

In addition to the ranking of pockets and the geometric/clustering and chemical features needed to derive it, SiteFerret returns to the user a significant amount of extra information.  
For instance, for a given entrance (as defined previously), 
the normal (with respect to the plane described by the protein atoms tangent to the probes) is returned (see \cref{fig:output}.a) together with the corresponding residues.
Similarly, for each bottleneck, SiteFerret returns to the user its axis and the list of associated residues.
The user can also access the entire list of normal vectors and residues associated with each pseudo-mouth, represented by aligned probes as described in \cref{sec:algorithm}.
SiteFerret also produces visual representations, which can be loaded in VMD\cite{Humphrey1996}. Pockets are represented in three ways: (i) as the actual cluster of probe spheres (\cref{fig:output}.a); (ii) in terms of tangent protein atoms (\cref{fig:output}.a and \cref{fig:output}.b, orange spheres); and (iii) as a triangulated \enquote{mold} of the clustered spheres (\cref{fig:output}.b), similarly to the representation of the pockets returned by NanoShaper. The representation (ii) is the most useful because it is explicitly related to the protein, and it is the one considered when applying the evaluation metrics described in \cref{sec:metrics} and used to build the tables discussed in the Results section. 

\section{Results}

As described in more detail in Sect.\ref{rank_protocols}, SiteFerret was directly compared against Fpocket and NS-Volume on the BM test set, 
on the LP dataset (holo and apo separately) and on a set of peptide-protein complexes excerpted from the BM dataset. 
Moreover, it was also compared to DeepSurf on the BM test set.

\subsection{Binding MOAD dataset}

Results are summarized in \cref{tab:moad}. While SiteFerret had a lower hit rate in the Top1 and Top3 positions, it outperformed NS-Volume and Fpocket on Top10. In analogy with previous findings\cite{Gagliardi2022}, DeepSurf was significantly better than the other methods. However, SiteFerret and DeepSurf had similar Top10 performances. This suggests that a successful strategy could involve directly learning the binding regions on the SES, rather than building the pockets first and then using descriptors for ranking.
As indicated in the table footnote, about $87\%$ of the top-ranked 3 subpockets by SiteFerret within successful parent pockets also matched a binding site. 
This suggests the analysis proposed in the lower part of the table, where the PC requirement is increased from $20\%$ to $50\%$. This stricter threshold limits the portion of a successful pocket that is not in contact with the ligand. As observed in the table, this new requirement decreases the performance of all methods. However, adopting the Nested Evaluation in SiteFerret strongly reduces the gaps with NS-Volume and Fpocket on the top-ranking positions, with SiteFerret surpassing Fpocket in Top3, and SiteFerret being superior to all methods, including DeepSurf, on Top10.
However, for the single-pocket evaluation method, where the master pockets are substituted by their three best-ranking subpockets (if any), SiteFerret's performance decreases significantly. This is because some of the largest master pockets are good matches for large ligands, in contrast to their subpockets. It is also due to the significant increase in the number of pockets that are put into the "hit parade" when the master pockets are unpacked, which push some successful subpockets beyond the 10 position.
Finally, NS-Volume clearly outperforms Fpocket when increasing the Pocket Coverage, showing the value of both the generation of smaller pockets and a simple volume-based ranking criterion.

\begin{table}
	\centering
	\caption{SiteFerret performance against Fpocket, NS-Volume, and DeepSurf in a random subsection of the binding MOAD dataset (total: 229 binding sites), Data for DeepSurf, NS-Volume, and Fpocket are obtained from unpublished data in Ref.\cite{Gagliardi2022}. DeepSurf hyperparameters are specifically optimized for the binding MOAD dataset. We also report on the average LC and PC scores of successful pockets.
	\label{tab:moad}}
\begin{tabular}{l|c|c|c|c|c|}
	\toprule
	Algorithm  & Top1   & Top3   & Top10 & LC& PC \\
	\midrule
	\multicolumn{4}{c}{\small Evaluation Thresholds: $LC\geq 50\%$, $PC\geq 20\%$}	\\
	\midrule
	Fpocket    & $62.9 $ & $77.7 $ & $87.3$ & $93.7$ & $64.3$\\
	NS-Volume  & $61.1$ & $79.5$ & $86.5$ & $89.3$ & $74.4$ \\ 
	SiteFerret\textsuperscript{\emph{a}} & $41.1$ & $67.7$ & $88.2$ & $94.8$ & $59.0$\\ 
	DeepSurf & $91.3$ & $92.6$ & $92.6$ & $95.6$ & $67.3$ \\ 
	\midrule
	\multicolumn{4}{c}{\small Evaluation Thresholds: $LC \geq 50\%$, $PC \geq 50\%$}	\\
	\midrule
	Fpocket    & $41.9$ & $55.0$ & $62.5$ & $94.7$ & $77.8$ \\ 
	NS-Volume  & $46.7$ & $63.8$ & $71.2$ & $89.5$ & $83.9$\\
	SiteFerret - \textsc{nested evaluation}\textsuperscript{\emph{b}} & $36.7$ & $59.8$ & $79.9$ & $92.6$ & $75.3$\\
	SiteFerret - \textsc{single pocket evaluation} & $24.5$ & $41.5$ & $58.5$ & $79.9$ & $89.1$\\
	DeepSurf & $69.4$ & $71.6$ & $71.6$ & $95.9$ & $76.0$ \\
	\bottomrule
\end{tabular}		

{\scriptsize\textsuperscript{\emph{a}} $87.1\%$ of Top3 subpockets hit the ligands. If the nested evaluation was adopted, all hit scores would increase by about $3\%$.}

{\scriptsize\textsuperscript{\emph{b}} $16.7\%$ of hits were possible thanks to 3 top-ranked subpockets.}
\end{table}

\subsection{LIGSITE-PocketPicker database}
\label{sec:LP}
As anticipated in \cref{sec:database}, we analyzed this database due to its large popularity in protein-ligand benchmarks. Importantly, in addition to the standard bound complexes, corresponding apo (unbound) structures are also present. This is important because, in any realistic application of a binding site predictor, only apo structures would be available, which could significantly differ from their holo counterparts due to the rearrangements induced by the binding\cite{Surade2012,Amaro2019}.

%

\paragraph{Holo}

The 48 bound structures of the LIGSITE-PocketPicker (LP) database, after the filtering steps described in \cref{sec:database}, 
comprise 57 binding sites.
Results are shown in \cref{tab:PP} for SiteFerret against Fpocket and NS-Volume. 
Similarly to what we observed for the BM dataset, SiteFerret's performance on the Top1 and Top3 ranked pockets is inferior to Fpocket and NS-Volume, and superior on the Top10. 
Again, the second half of the table shows that, when the stricter requirements apply, SiteFerret outperforms the competitors in the Nested Evaluation protocol, while it is inferior when the Single Pocket protocol is used.

\begin{table}
	\centering
	\caption{Performance against Fpocket and NS-Volume on the LP bound database.
		\label{tab:PP}}
	\begin{tabular}{l|c|c|c|c|c}
		\toprule
		Algorithm  & Top1   & Top3   & Top10 & LC & PC\\
		\midrule
		\multicolumn{4}{c}{\small Score Thresholds: $LC \geq 50\%$, $PC \geq 20\%$}\\
		\midrule
		Fpocket    & $68.4$ & $89.5$ & $93.0$ & $96.3$ & $63.9$\\
		NS-Volume   & $68.4$ & $84.2$ & $91.2$ &$94.7$ & $83.6$\\
		SiteFerret \textsuperscript{\emph{a}} & $64.9$ & $77.2$ & $94.7$ &$97.0$& $60.5$  \\
		\bottomrule
		\multicolumn{4}{c}{\small Score Thresholds: $LC \geq 50\%$, $PC \geq 50\%$}\\
		\bottomrule
		Fpocket   & $50.9$ & $73.7$ & $73.7$ & $96.8$ & $72.7$ \\
		NS-Volume   & $59.7$ & $75.4$ & $82.5$ & $94.7$ & $88.7$ \\
		SiteFerret - \textsc{nested evaluation} \textsuperscript{\emph{b}} & $63.2$ & $75.4$ & $93.0$ & $94.6$& $75.6$ \\
		SiteFerret - \textsc{single pocket evaluation} & $36.8$ & $45.6$ & $64.9$ & $85.9$ & $86.9$   \\
		\bottomrule
	\end{tabular}	

{\scriptsize\textsuperscript{\emph{a}} $92.6\%$ of top3 subpockets hit the ligands. The nested evaluation here would not bring benefits (all master pockets hit without \enquote{help}) from their subpockets.} 

{\scriptsize\textsuperscript{\emph{b}} $21.1\%$ of hits were realized thanks to subpockets matching.}
\end{table}

\paragraph{Apo}

The literature does not define well the procedure to assess the performance of pocket retrieval methods over unbound structures.
Since the exact position of the ligand in the apo form is unknown by definition, a specific procedure was devised to map binding residues in the holo structure to the apo form. This procedure is detailed in \cref{sec:database}.  
The LC and PC score definitions are preserved, provided that one refers to residues and residue numbers rather than atoms. 
Notably, by slightly increasing the probe radius used to identify exposed residues via the SES, one can strongly impact Fpocket's performance, which tends to include buried residues in its pockets. This is shown in detail in the Supplementary Material.

In terms of residues, the cardinality of the clusters is much smaller. As such, a quantitative comparison of the LC and PC score thresholds with the previous analysis based on atoms is not obvious.
Here, we consider the same thresholds as used before for LC and PC, as well as a smaller LC threshold of $20\%$.
As illustrated in \cref{tab:PP_apo}, the results are more sensitive to the adopted thresholds.
Finally, we note that NS-Volume's performance is significantly worse than Fpocket and SiteFerret for most of the evaluation measures. This seems to be due to NS-Volume's tendency to present small pockets, with too few exposed residues. Indeed, NS-Volume's performance significantly improved when we lowered the threshold on the LC score to $20\%$.

\begin{table}[h]
	\centering
	\caption{Performance on the LP unbound (apo) database.
		\label{tab:PP_apo}}
	\begin{tabular}{l|c|c|c|c|c}
		\toprule
		Algorithm  & Top1   & Top3   & Top10 & LC & PC\\
		\midrule
		\multicolumn{4}{c}{\small Score Thresholds: $LC \geq 50\%$, $PC \geq 20\%$ on residues}	\\
		\midrule
		Fpocket    & $49.1$ & $78.2$ & $83.6$ & 75.1 & 70.4\\
		NS-Volume   & $58.2$ & $63.6$ & $65.5$ & 79.9 & 78.6\\
		SiteFerret  & $50.9$ & $76.4$ & $90.9$ & 80.2 & 68.0\\
		\midrule
		\multicolumn{4}{c}{\small Score Thresholds: $LC \geq 50\%$, $PC \geq 50\%$ on residues}\\
		\bottomrule
		Fpocket    & $41.8$ & $70.9$ & $74.5$ & 74.8 & 74.9 \\
		NS-Volume   & $50.9$ & $58.2$ & $60.0$ & 79.9 & 82.4 \\
		SiteFerret, \textsc{nested evaluation}\textsuperscript{\emph{a}} & $40.0$ & $63.6$ & $78.2$ & 80.8 & 73.9  \\
		SiteFerret, \textsc{single pocket evaluation} & $10.9$ & $29.1$ & $43.6$ & 56.8 & 93.6 \\
		\bottomrule
		\multicolumn{4}{c}{\small Score Thresholds: $LC \geq 20\%$, $PC \geq 50\%$ on residues}	\\
		\bottomrule
		Fpocket    & $43.6$ & $74.5$ & $80.0$ & 71.5 & 72.2 \\
		NS-Volume   & $56.4$ & $72.7$ & $80.0$ & 67.9 & 81.7\\
		SiteFerret, \textsc{nested evaluation}\textsuperscript{\emph{b}}
		&  $49.1$ & $74.5$ & $92.7$ & 70.7& 70.2\\
		SiteFerret, \textsc{single pocket evaluation} &  $38.2$ & $70.9$ & $92.7$ & 39.2 & 79.4 \\
		\bottomrule
	\end{tabular}	

{\scriptsize\textsuperscript{\emph{a}} $0\%$ of hits were realized thanks to subpockets matching, so \emph{nested evaluation} does not apply.\\
\textsuperscript{\emph{b}} $12.7\%$ of hits were realized thanks to Top3 subpockets matching.}
\end{table}

\subsection{Protein-Peptide binding sites}
\label{sec:results:peptides}

Finally, we show the results for the set of 115 protein-peptide binding sites.
Peptides are larger than ligands and tend to occupy large riverbed-like pockets.
It is interesting to assess the method's performance on protein-peptide sites since they are both geometrically and chemically similar to protein-protein interaction sites. The prediction of protein-protein interaction sites is distinct from and  more complex than protein-ligand binding prediction and goes beyond the scope of the present work. However, there is remarkable evidence that a computational method can cover both tasks.

As illustrated in \cref{tab:peptides}, SiteFerret is significantly better on this database than Fpocket and NS-Volume on all figures of merit.
Here, the number of subpocket hits decreases from about $90\%$, found previously on ligand/small molecule binding predictions, to $65\%$. This is due to the size of the peptides, which are often significantly larger than the subpockets.

\begin{table}
	\centering
	\caption{Performance on the BM subset containing peptide binding sites (115 sites). Scoring thresholds: $LC\geq50\%$ and $PC\geq 20\%$.
		\label{tab:peptides}}
	\begin{tabular}{l|c|c|c|c|c}
		\toprule
		Algorithm  & Top1   & Top3   & Top10 & LC & PC\\
		\midrule
		Fpocket   & $30.0$ & $49.6$ & $58.3$ & $77.9$ & $65.7$\\ 
		NS-Volume & $37.4$ & $49.6$ & $55.7$ & $77.7$ & $76.1$\\
		SiteFerret\textsuperscript{\emph{a}} & $37.4$ & $61.7$ & $86.1$ & $79.9 $ & $66.5 $ \\
		\bottomrule
	\end{tabular}	
	
	{\scriptsize\textsuperscript{\emph{a}} $65\%$ of top3 subpockets hit the peptide. If the nested evaluation was adopted, all hit scores would increase by about $1.5\%$ }
\end{table}

\subsection{Subpocket identification and characterization}
\label{sub-pocket_analysis}
To our knowledge, only a few methods leverage the concept of subpockets: DogSite (and similarly Lsite and Dsite, discussed by the same authors)\cite{Volkamer2010} and the more recent CAVIAR method\cite{Milanetti2021}. In contrast to CAVIAR, which ranks with an ad hoc scoring based on size and \enquote{buriedeness}, DogSite uses a machine-learned druggability score (DogSiteScorer)\cite{Volkamer2012}. However,
a criticism of DogSite is its tendency to generate very large subpockets, which do not capture small and localized parts of cavities that potentially encompass defined functional groups of ligands. This ability is instead attributed to CAVIAR\cite{Milanetti2021}.
These methods could not be directly included in the current performance analysis due to the different figures of merit adopted to evaluate a good hitting pocket, which is based on a geometric center criterion in one case (a pocket is successful if its geometric center lies within $4\AA$ of \emph{any} ligand atoms, DogSite) and based on a weak single ligand atom overlapping criterion in the other case (it is sufficient that one ligand atom overlaps a pocket grid point for the latter to be considered a correct binding site, CAVIAR). However, we provide below a qualitative analysis supported by some examples. 
We start by analysing the same examples discussed in the CAVIAR paper, which also provided a critical comparison with DogSite. In general, SiteFerret performs a subpocket segmentation that more closely resembles the segmentation obtained by CAVIAR rather than that of DogSite. We then provide examples of shallow sites where SiteFerret outperforms several methods, including CAVIAR and Fpocket. 

\begin{figure}[h!]
\caption{Chaperone protein hsp90-$\alpha$ (PDB code 2FWZ) co-crystallized with the water soluble inhibitor PU-H71. Red and orange: first and second top-ranked subpockets of the top-ranked pocket generated by SiteFerret. \label{fig:2fwz}}
	\includegraphics[width=0.5\linewidth]{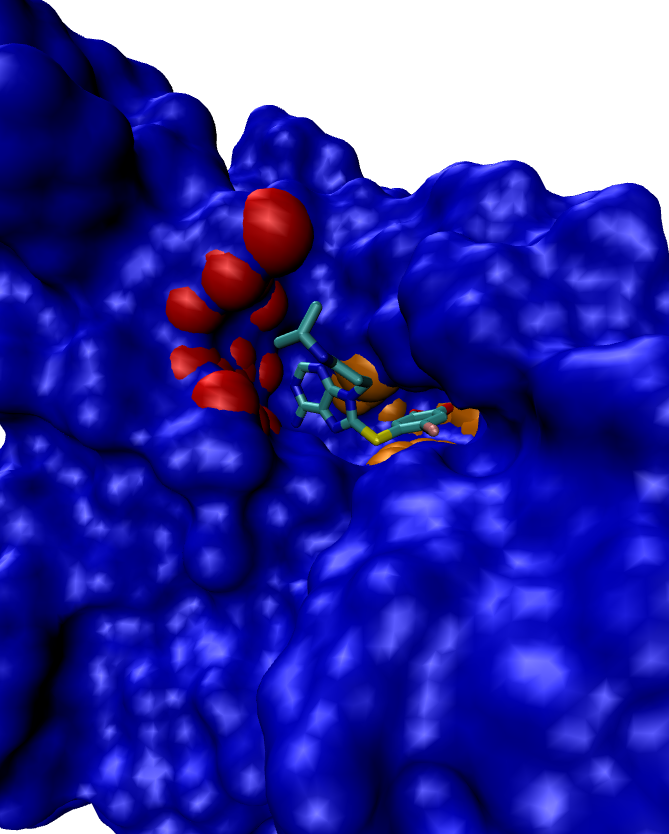}
\end{figure}

\paragraph{HSP90-$\alpha$.} The first example is the binding pocket of the chaperone protein HSP90-$\alpha$, PDB code 2FWZ.
SiteFerret topranks a large master pocket containing three subpockets. As shown in \cref{fig:2fwz}, the first-ranked subpocket (red) is occupied by the ligand's adenine head group, while the second-ranked subpocket (orange) is occupied by the ligand's iodo-benzodioxole group. As shown in Ref.\cite{Milanetti2021}, this result is very similar to what was obtained by CAVIAR, while DogSite overspans the two subpockets.

\begin{figure}[h!]
\caption{SiteFerret result for HIV-1 protease  (PDB code 1C70) co-crystallized with the inhibitor L-756423. The channel where the ligand resides is identified by the top-ranked pocket.
 Left panel: one entrance to the channel containing the ligand. Top-ranked subpocket in the foreground (red). Two other subpockets visible deeper in the channel. Right panel: second entrance to the channel highlighted by the 2nd (orange) and 4th (green) ranked subpockets.
		\label{fig:1c70}}
	\begin{subfigure}{0.42\linewidth}
		\includegraphics[width=\linewidth]{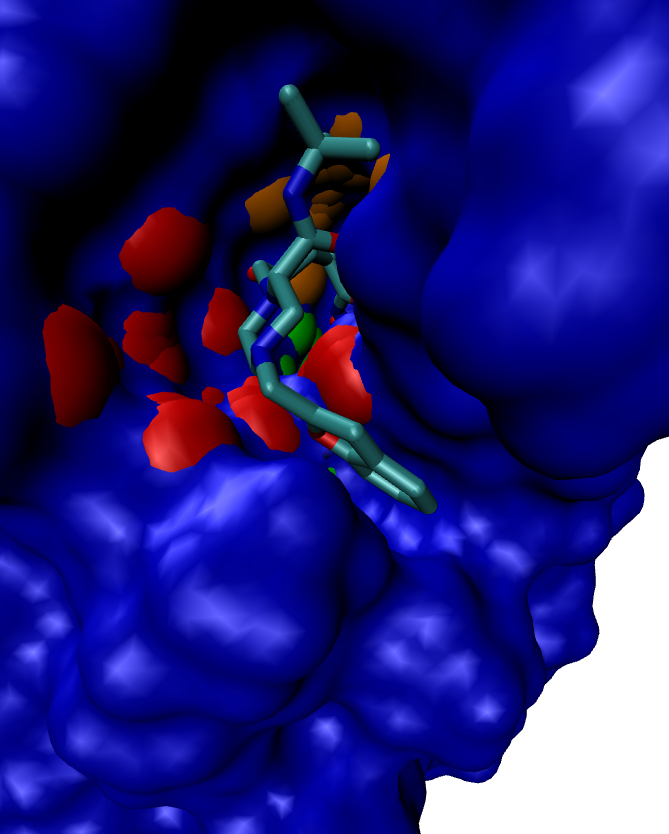}
	\end{subfigure}
\begin{subfigure}{0.42\linewidth}
	\includegraphics[width=\linewidth]{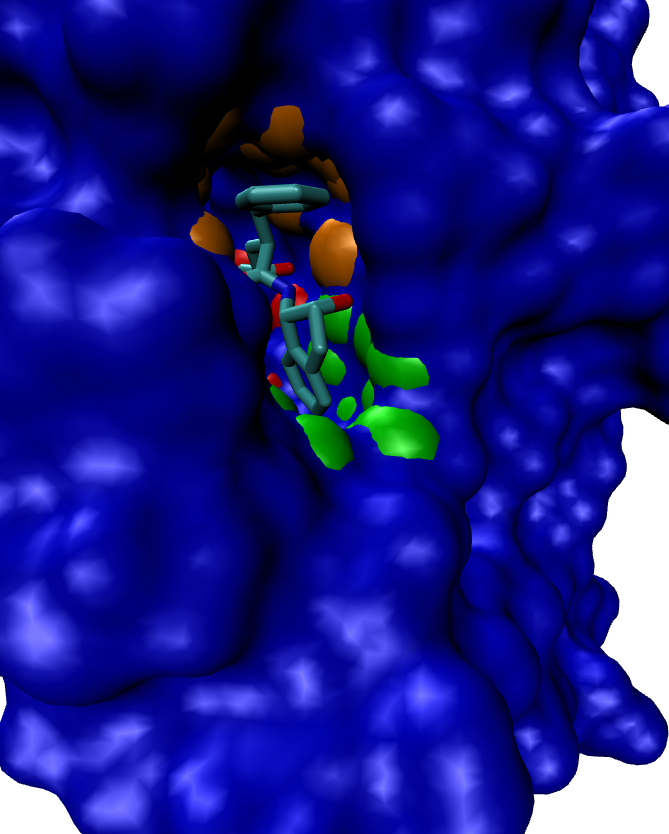}
\end{subfigure}
\end{figure}

\paragraph{HIV-1 protease.} Another example is that of HIV-1 protease (PDB code 1C70), where CAVIAR identifies seven subcavities, six of which correspond to the standard subsites of specific amino acid side chains of the peptidic substrate\cite{Milanetti2021}. In this case, DogSite outputs only a single pocket. Our result lies somewhat between the results of CAVIAR and DogSite. Indeed, our top-ranked pocket contains 3 out of 5 subpockets corresponding to sites occupied by the ligand. This is shown in \cref{fig:1c70}, where the ligand occupies a channel and the top-ranked subpocket (red) points to one of the cavity entrances. The 2nd and 4th ranked subpockets (orange and green) describe two dips within the wide second entrance to the channel.

\begin{figure}
\caption{HCV NS3 protease co-crystallized with Simeprecir (PDB code 3KEE, only chain A). Left panel: SiteFerret result; white, whole pocket (second-ranked); red and orange, first and second top-ranked subpockets. 
Right panel, Fpocket result (pink patch). \label{fig:3kee}}
    \begin{subfigure}{0.45\linewidth}
        \includegraphics[width=\linewidth]{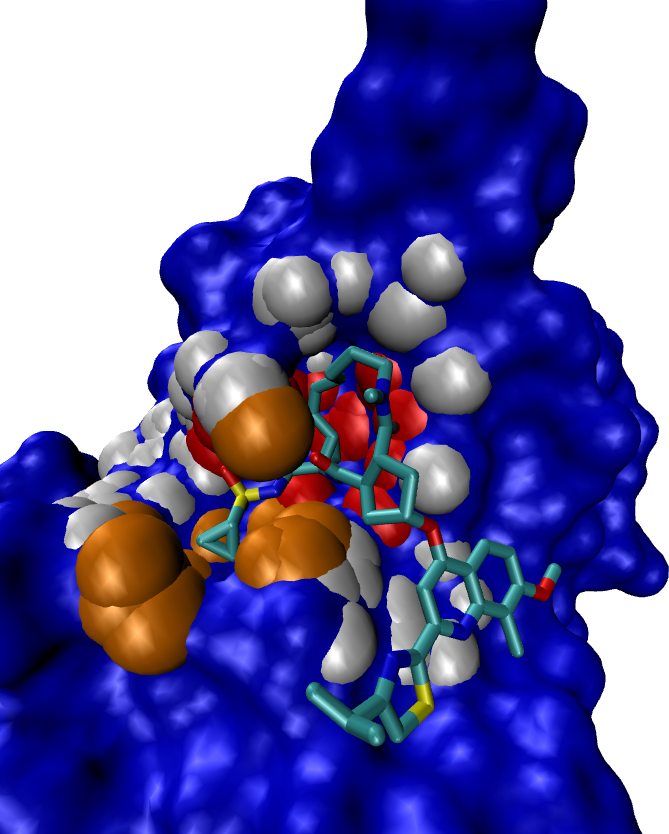}
    \end{subfigure}
    \begin{subfigure}{0.45\linewidth}
    	\includegraphics[width=\linewidth]{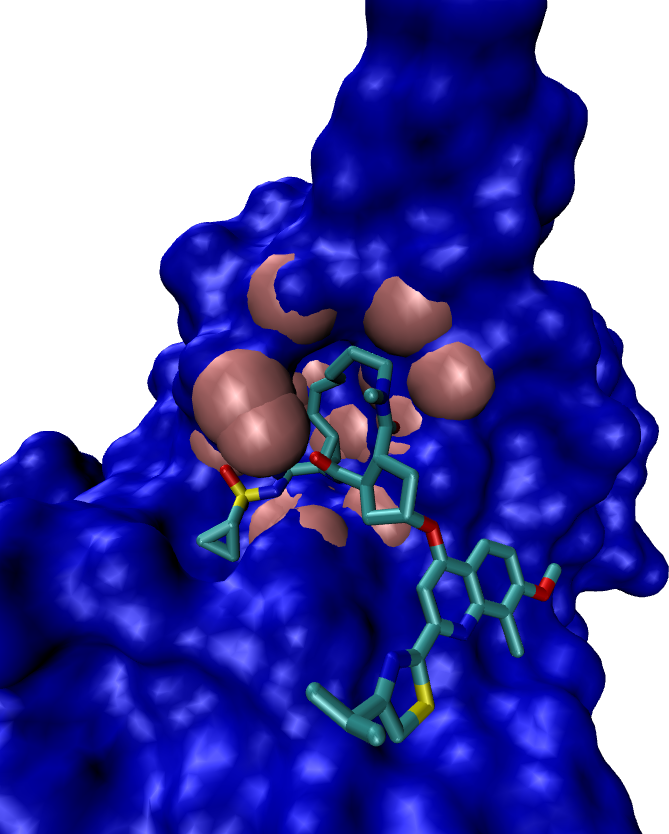}
    \end{subfigure}
\end{figure}

\paragraph{HCV NS3 protease.} A further interesting example is HCV NS3 protease (PDB code 3KEE). The binding pocket, ranked second, is illustrated in \cref{fig:3kee} (white). The part of the binding site corresponding to the deeper groove is segmented into 2 subpockets (red, ranked first, and orange, ranked second) describing cavities separated by a cleft (superficial bottleneck). These subpockets are similar to the one reported by CAVIAR.
However, while CAVIAR and DogSite do not include the remaining relatively shallow region where the ligand binds (figs.~4c and d in Ref.\cite{Milanetti2021}), this is part of the master pocket in SiteFerret (white). 
Similarly, as shown in \cref{fig:3kee}b, when running Fpocket (second-ranked pocket shown), the pocket partially matching the ligand (pink) roughly corresponds to the top-ranked subpocket in SiteFerret.
This last example also illustrates SiteFerret's ability to overcome the aforementioned difficulty in describing shallow sites, which is commonly found in other pocket detection algorithms.
Indeed, while its top-ranking performance is poorer than the best state-of-the-art techniques, our method performs outstandingly on difficult shallow sites. This is suggested by the results on peptide binding sites \cref{tab:peptides} and can be further illustrated with the following examples.

\begin{figure}
	\caption{The white patches represent the pockets returned by SiteFerret. First panel: HIV-integrase (PDB code 3LPT), 8th ranked pocket. Second panel: Hexameric insulin (PDB code 1MPJ), 5th ranked. Third panel: Complex between methyl $\alpha$-D-mannopyranoside and concanavalin A (PDB code 5CNA, restrained to chain A), 8th ranked. 
		\label{fig:shallow}}
	\begin{subfigure}{0.3\linewidth}
		\includegraphics[width=\linewidth]{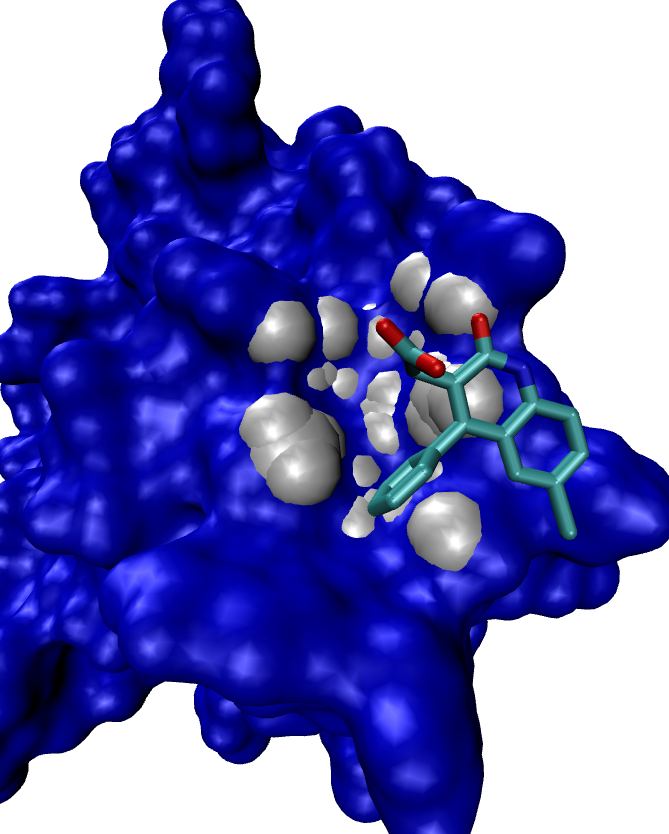}
	\end{subfigure}
	\begin{subfigure}{0.3\linewidth}
		\includegraphics[width=\linewidth]{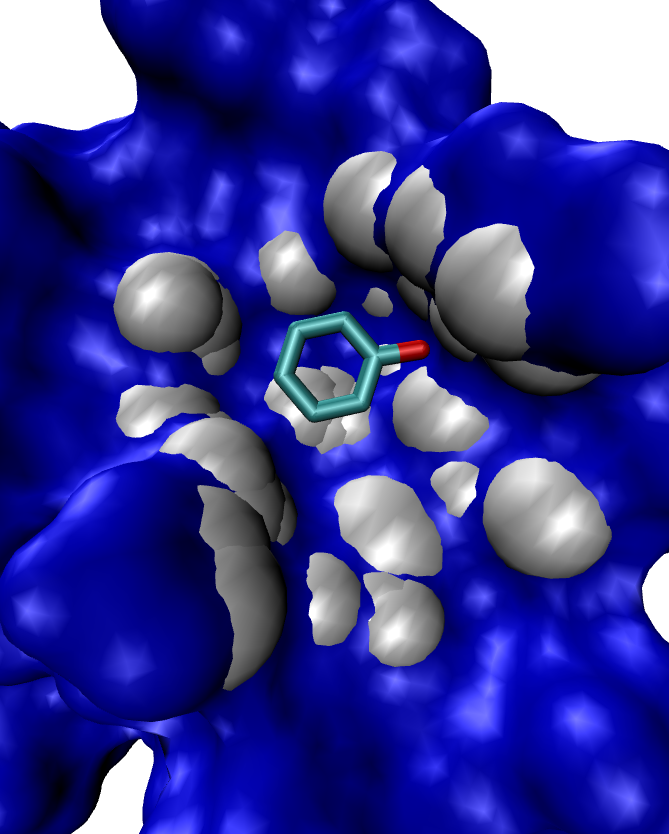}
	\end{subfigure}
\begin{subfigure}{0.3\linewidth}
	\includegraphics[width=\linewidth]{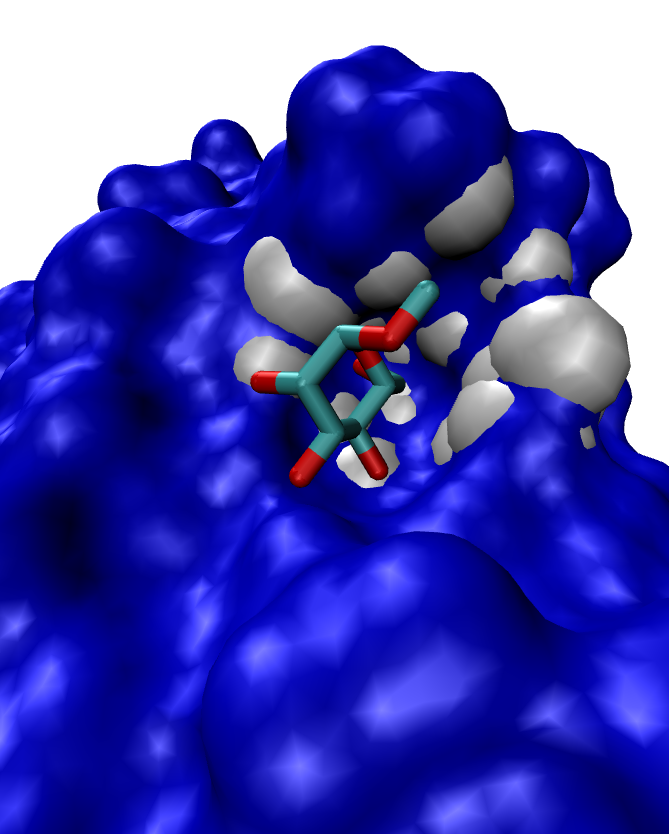}
\end{subfigure}
\end{figure}

\paragraph{HIV-integrase, Hexameric insulin, and a few systems from the LP database.} Let us now consider HIV-integrase, PDB code 3LPT \cite{Christ2010}, 
a structure where both DogSite and Fpocket fail to identify the observed binding site\cite{Volkamer2012}. As illustrated in \cref{fig:shallow}a, SiteFerret reports this site as the 8th-ranked pocket.
Another notably hard example is Hexameric insulin with its ligand methylparaben\cite{Whittingham1995}, PDB code 1MPJ, where, as reported in Ref.\cite{Huang2006} LIGSITE$^{csc}$, LIGSITE, PASS, CAST, and SURFNET fail. In addition to those methods, we tested Fpocket and found that it could not correctly identify the binding pocket. As illustrated in \cref{fig:shallow}.b, SiteFerret predicted the correct binding pocket, which ranks 5th.
Finally, SiteFerret reproduced two of the binding sites in the LP dataset (PDB codes 3MTH and 5CNA), where both Fpocket and CAVIAR failed. These binding sites are characterized by a very exposed and flat surface of interaction.
PDB code 3MTH corresponds to a conformation of Hexameric insulin with methylparaben\cite{Whittingham1995}; here, the binding pocket ranks in the 5th position (not shown). 
PDB code 5CNA corresponds to the complex between methyl $\alpha$-D-mannopyranoside and concanavalin A. Restricting our analysis to chain A, as in the LP database, SiteFerret correctly identifies the binding pocket, and places it at the 8th position. This is shown in \cref{fig:shallow}.c. 

\subsection{SHAP importance analysis of features}
\begin{figure}[h]
	\centering
	\caption{SHAP importance analysis of the features used by 
 the Isolation Forests. Every point corresponds to a sample. 
 The color map, from blue to red, runs from the smallest to the largest value for the given feature.
The rightmost points count towards being deemed inlayer samples (i.e. ligandable pockets). The leftmost points correspond to feature values counting towards being outlayer. On the left panel: geometric and clustering features. On the right panel: chemical features (20 residues and hydrophilicity scores)\cite{Lundberg2020}. 
	\label{fig:shap}}
	\includegraphics[width=0.9\linewidth]{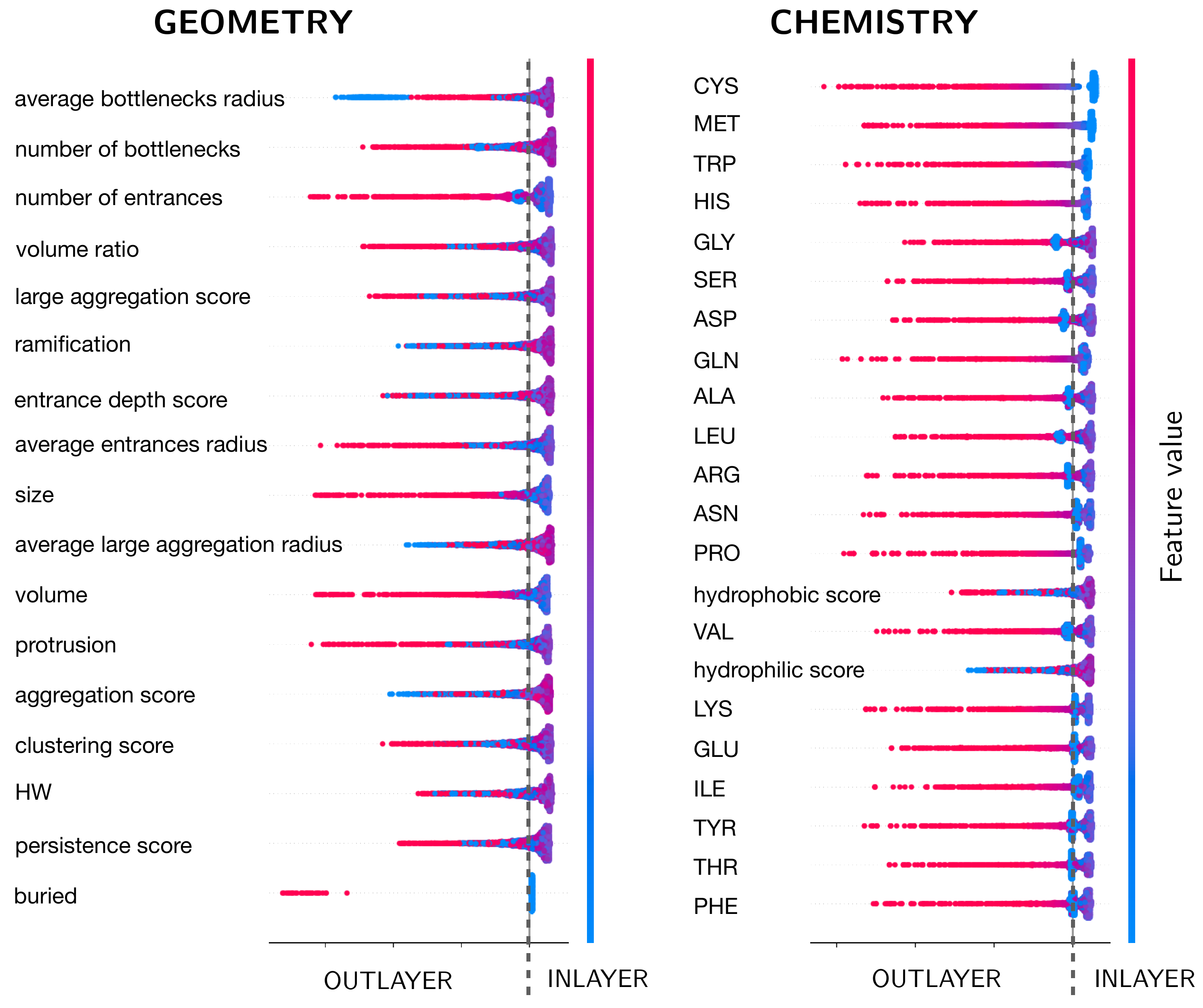}
\end{figure}

The SHAP importance analysis is shown in \cref{fig:shap} for the clustering/geometric IF (left panel) and the chemical IF (right panel).

The analysis is done on the training database and it highlights the most representative features of binding pockets observed by SiteFerret in this work.
On the y-axis, the plot sorts the features by importance. On the x-axis, the plot represents the SHAP value i.e. the relative impact on the model outcome. Values on the left of the dashed vertical line (also called the \enquote{decision} line) point to an "anomaly", i.e. to a nonbinding site, while the values on the right of the dashed vertical line point to a "normal" point (i.e. a plausible binding pocket).

Finally, each dot represents a realization of a random subset of the training set. As illustrated in the plot, the numerical values of each feature are color-coded from blue (lower values) to red (higher values).

\paragraph{Geometric and clustering descriptors:} The topmost position is occupied by the average radius associated with bottlenecks and the number of bottlenecks. Given the feature values, it then appears that the presence of bottlenecks counts positively, but that too many bottlenecks constitute a strong anomaly score (red dots with large negative SHAP value). This scenario likely corresponds to very large pockets. Similarly, many entrances and/or a large effective radius of the entrance count towards an anomaly and highly impact the score. Many entrances are likely correlated to large canyon-like superficial pockets. 
Moreover, features ranked high in importance include features related to complexity, size of the branching, and other features related to the clustering process. Indeed, \emph{ramification}, \emph{large aggregation score}, \emph{entrance depth score}, and \emph{large-aggregation radius} all appear to be highly important. The tendency of druggable pockets to be more complex in shape was also reported in\cite{Volkamer2012}. In contrast, a tendency to partially favor more compact pockets is shown by the \emph{volume ratio} (high in ranking) together with other less relevant descriptors related to geometric compactness and size. The lower importance of these descriptors could also be due to a partial redundancy with the others.

 \paragraph{Chemical descriptors:} 

The SHAP analysis of the chemical IF shows residues such as CYS, MET, and TRP at the topmost positions, while some appear low in importance (e.g. PHE). As discussed in \cref{sec:discussion}, this result has some interesting analogies with previously published broader statistical analysis of the binding MOAD dataset\cite{Khazanov2013}.

Concerning the hydrophilic and hydrophobic scores, our findings confirm that ligandable pockets mostly tend to be hydrophilic\cite{Volkamer2012} (as shown by \cref{fig:shap}, both large hydrophobic score values and low hydrophilic score values count against being a plausible binding site). However, we also show that a certain degree of hydrophobicity is acceptable (purplish color on the right-hand side of the decision line, while excessive hydrophobic values count as an anomaly\cite{LeGuilloux2009}).
The SHAP analysis also suggests that the hydrophobic score is a stronger indicator than the hydrophilic score. 
As shown in Fig.9 in the Supplementary Material, this is justified by the distribution shape, which is narrower for the hydrophobic score than the hydrophilic score (see also \cref{sec:discussion}).

\section{Discussion}
\label{sec:discussion}

The identification of potential binding sites in a protein structure is a relevant and much-studied problem. In this context, it is interesting to identify the features that make a pocket a plausible binding site. Furthermore, as pointed out by Simoes et.~al\cite{Simoes2017}, the current problems with site detection algorithms include the capacity to produce a meaningful hierarchical segmentation of cavities (expressed here with the subpockets), and the need to return both standard deep clefts and shallow sites (grooves). The latter task often requires dedicated procedures (e.g.~in sphere-based methods to distinguish shallow sites from deeper invaginations, it could be critical to correctly map the size of the "entrance" to the value of the mouth radius\cite{Tripathi2010}).

As shown here, SiteFerret is remarkably successful in reporting, among the top ten positions, pockets that significantly overlap with the observed binding site(s), either as a whole or via one of their three main subpockets. The significance of the subpocket segmentation is reflected in the remarkable performance achieved when stricter requirements are adopted in terms of PC.

Globally, evidence suggests that SiteFerret can correctly identify shallow sites, although they are rarely ranked in the highest positions. This is expected since these sites are less frequent and thus less represented in the training set.

\paragraph{Feature importance analysis.} In summary, a good binding pocket is generally determined by a combination of significant clustering complexity and some geometric compactness, especially concerning superficial features (e.g.~entrance number and radius, and \emph{protrusion}). A certain degree of complexity in the pocket's structure also seems desirable (i.e. bottlenecks, \emph{ramification}, entrance depth). 
Interestingly, the analysis of chemical features mostly confirms the previous analysis of the BM database in Ref.~\cite{Khazanov2013}. 
That study compared residues in contact with ligands with respect to the rest of the protein surface, and considered both valid and invalid sites (according to the bindingMOAD definition). The study concluded that the following residues are preferred indicators of a pocket's ligandability: ILE, MET, PHE, CYS, GLY, TYR, TRP. 
The SHAP analysis of the IF reached the same conclusion about the relative residue population of CYS, MET, TRP, GLY, which all appear in the topmost positions. Conversely, ILE, TYR, and PHE are found in the lowest positions of the graph. This strong agreement is surprising, given that the IF has no information about the distribution of amino acids on the rest of the protein surface. As extensively discussed in the Supplementary Material, the feature importance can then only correlate with the shape of the observed distribution.
As shown in Fig.7 of the Supplementary Material, narrow unimodal distributions of relative residue populations will tend to be placed on the topmost positions (e.g.~CYS and MET), while residues characterized by a bimodal wide distribution are not a strong anomaly indicator (e.g. PHE).
In the latter case, since many valid pockets have a wide variety of admissible values, this is not discriminant when comparing potential pockets against each other.

\paragraph{IF discrimination ability.} In this work, we trained IFs on geometric/clustering and on compositional/chemical features, both individually and after merging the feature sets. Taken separately, the trained forests already have similarly high discrimination abilities (with chemistry being slightly better in the Top3 category). However, their combination, via averaging, improves the ranking performance, confirming that the information leveraged by the two forests is at least partially complementary.
This result is also slightly better (about $2\%$) than what would be obtained by considering a single IF trained on both geometric and chemical descriptors together. Finally, as also shown in the Supplementary Material, the IF ranking is superior to the simpler standard volume-based ranking. However, the difference is not great, confirming that a simple volume ranking remains a quick and reasonable strategy if coupled with a suitable pocket generation stage, when more refined approaches are not available\cite{Gagliardi2022}. 
In principle, the method can be tailored to a given dataset of interest. In this work, we chose the BM dataset as our reference for training and parameter optimization since it is fairly large and representative.
In conclusion, the IF's discrimination power was effective since, as shown in the Supplementary Material, we decided not to prune the set of pockets generated in the first stage, which leads to large numbers for a given structure (around 100 on average).

\paragraph{Comparison with other site predictors.} 
As described above, our main benchmark was Fpocket. Direct comparisons with other methods were prevented by: proprietary software; unclear normalization and ranking procedure in the reference papers; impossibility of applying the LC- and PC-based scoring system. Moreover, in many cases, other tools have already been compared against Fpocket, which is certainly one of the most competitive open-source pocket detection softwares available today. The possibility of using the same metric prompted us to include NanoShaper with a volume-based ranking criterion. 
Using standard assessment criteria, both methods outperformed SiteFerret on Top1 and Top3, while the latter had a better Top10 score. However, when there were stricter requirements in terms of pocket coverage, SiteFerret generally outperformed the competitors. Interestingly, as highlighted in Ref.\cite{Gagliardi2022}, pockets generated by NS with a simple volume-based ranking remain very competitive (see table 1 and 3) against our and other more complex scoring approaches.
Globally, DeepSurf, a tool based on deep neural networks, performs outstandingly in terms of providing a few putative binding sites. Its Top1 and Top3 were significantly higher than the other methods.

A comment is needed on the ability to identify and characterize subpockets. When considering the quality of the generated subpockets in several examples (see Sect. \ref{sub-pocket_analysis}), SiteFerret compares similarly to the recent CAVIAR method, which turns out to be superior to DogSite, at least as far as the sub-pocket segmentation is concerned\cite{Marchand2021,Volkamer2010}. To our knowledge, these are the only other tools leveraging the concept of subpockets, although they use a different approach from ours, as discussed above.

Interestingly, when considering the performance on protein-peptide binding sites, SiteFerret definitely outperformed Fpocket and NS-Volume in all the considered metrics.
Moreover, the results suggest that SiteFerret also performs well on shallow sites, while retaining a good performance on the most commonly found deep groove/invagination binding sites (typical of small binding molecules). 

\section{Conclusions}

In this work, we present SiteFerret, a method for identifying ligandable sites on protein structures. SiteFerret combines a pocket generation stage based on an ad hoc clustering method of SES probe spheres at different radii with the IF anomaly detector. 
Overall, SiteFerret was excellent in reporting observed binding sites among its Top10 results, while other approaches were superior in the Top1 or Top3 categories. SiteFerret had an unprecedented flexibility in identifying relevant binding regions in a wide variety of systems, while requiring minimal parametrization. It also reliably segmented putative binding pockets into smaller subpockets. Moreover, it effectively recognized protein-peptide binding regions and, more importantly, binding sites in unbound (apo) structures, as shown on the LigSite/PocketPicker database. Finally, SiteFerret provides a remarkable amount of descriptive information about the pockets.
When considering protein-peptide sites, SiteFerret is superior in all figures of merit to the other geometry-based approaches considered (Fpocket and NS-Volume). 
When stricter requirements are enforced in terms of Pocket Coverage score, which assesses a found site's capacity to more precisely pinpoint the binding region, SiteFerret is more robust than other methods, including state-of-the-art deep-neural-network-based approaches such as DeepSurf. This is thanks to the generated subpockets, which can more precisely fit the actual binding region.
In this work, we both qualitatively and quantitatively demonstrate the effectiveness of subpocket segmentation. Our method achieved a hit rate within successful (master) pockets of around $90\%$ in most of the databases considered, even though the analysis was limited to only the first three subpockets of each master.
Finally, we also demonstrate SiteFerret's outstanding ability to detect difficult shallow binding sites. 

On the technical side, this work also shows that one can use a standard anomaly detector such as the IF as a one-class classifier, and that it can be used to combine the effects of geometric/clustering and chemical descriptors. Furthermore, we confirm that the commonly used volume-based ranking remains a competitive strategy.

Given these results and its flexibility, SiteFerret is a promising tool for bridging the gap between the task of recognising small-molecule binding sites and the task of recognising protein-peptide binding regions. This suggests the prediction of protein-protein interaction interfaces as a natural extension of this work.

\section*{Code availability}
The code is freely available at \url{https://github.com/concept-lab/SiteFerret}.
We also provide in the same repository scripts and tools used to generate the data shown in the Results section.
\begin{acknowledgement}
We thank Pedro Reis for performing the sequence alignment of apo residues and useful discussion.
We thank the IMATI-CNR group, and especially Andrea Raffo and Silvia Biasotti, for providing useful insights in the early stage of the project.
\end{acknowledgement}

\bibliography{pickPocket}

\end{document}